\documentclass[preprint,12pt]{elsarticle}

\usepackage[T1]{fontenc}
\usepackage[utf8]{inputenc} 
\usepackage[french, english]{babel}
\usepackage[cyr]{aeguill}

\usepackage{upgreek} 

\usepackage[colorlinks=wrong]{hyperref}

\usepackage{xspace}

\usepackage{lmodern} 

\usepackage[top=4.cm, bottom=4.cm, left=4.cm, right=3.cm,headheight=17pt,,includehead,includefoot,heightrounded]{geometry} 

\usepackage{emptypage} 

\usepackage{appendix}

\usepackage{subcaption} 
\usepackage{graphicx} 
\usepackage{caption} 
\usepackage{wrapfig} 
\usepackage{here} 

\usepackage{breakcites} 

\usepackage{float}
\usepackage{array}

\usepackage{tikz} 
\usepackage{tikz-cd}
\usetikzlibrary{arrows,calc,chains,matrix,positioning,scopes} 

\usepackage{tabularx}
\usepackage{multirow}
\usepackage{multicol}
\usepackage{bigdelim}

\usepackage{bbm} 

\usepackage{listings} 

\usepackage{latexsym}
\usepackage{amsthm} 
\usepackage{amsmath} 
\usepackage{amsfonts}
\usepackage{amsxtra}
\usepackage{yhmath}
\usepackage{amssymb} 
\usepackage{mathrsfs} 
\usepackage{esint} 
\usepackage{empheq}
\usepackage{stmaryrd}
\usepackage{blkarray} 

\usepackage{cleveref}
\crefformat{footnote}{#2\footnotemark[#1]#3}

\usepackage{cancel} 

\usepackage{keyval}

\usepackage{comment}

\newcommand{\abs}[1]{\left\lvert#1\right\rvert}

\newcommand{\determinant}[2][]{\mathrm{det}_{#1}\mathopen{}\left(#2\right)}

\newcommand{\enlacement}[3][]{\mathop{}\mathopen{}\mathrm{{\ell}k}_{#1}\left(#2,#3\right)}

\newcommand{\slfrac}[2]{\left.#1\middle/#2\right.}

\DeclareSymbolFont{UPM}{U}{eur}{m}{n}
\DeclareMathSymbol{\uppartial}{0}{UPM}{"40}

\newcommand{\ensemblenombre}[1]{\mathbb{#1}}
\newcommand{\N}{\ensemblenombre{N}}
\newcommand{\Z}{\ensemblenombre{Z}}
\newcommand{\Q}{\ensemblenombre{Q}}
\newcommand{\R}{\ensemblenombre{R}}
\newcommand{\Cplx}{\ensemblenombre{C}}

\newcommand{\RP}{\ensemblenombre{RP}}

\theoremstyle{definition}
\newtheorem{defi}{Definition}

\newtheorem{rema}{Remark}

\newtheorem{exple}{Example}
\newtheorem{thm}{Theorem}
\newtheorem{cons}{Consequence}
\newtheorem{coro}{Corollary}
\newtheorem{prop}{Proposition}

\makeatletter
\def\ps@pprintTitle{%
  \let\@oddhead\@empty
  \let\@evenhead\@empty
  \let\@oddfoot\@empty
  \let\@evenfoot\@oddfoot
}
\makeatother

 \usepackage{soul}
\begin{document}

	\begin{frontmatter}
		
		\title{$\mathrm{U}(1)^{n}$ Chern-Simons theory: partition function, reciprocity formula and Chern-Simons duality}
		
		\author[add1]{Han-Miru Kim}
		\ead{kimha@ethz.ch}
		\author[add2]{Philippe Mathieu}
		\ead{philippe.mathieu@math.uzh.ch}
		\author[add3]{Michail Tagaris}
		\ead{michail.tagaris@lapth.cnrs.fr}
		\author[add4]{Frank Thuillier}
		\ead{frank.thuillier@lapth.cnrs.fr}  
		
		\address[add1]{ETH, Zürich, Schweiz}
		\address[add2]{Institut für Mathematik, Universitât Zürich, Winterthurerstrasse 190, CH-8057 Zürich, Schweiz}
        \address[add3]{LAPTh, CNRS, USMB, 9 Chemin de Bellevue, 74940 Annecy, France}
		\address[add4]{LAPTh, CNRS, USMB, 9 Chemin de Bellevue, 74940 Annecy, France, (corresponding author)}
		
		\begin{abstract}
            The $\mathrm{U}(1)$ Chern-Simons theory can be extended to a topological $\mathrm{U}(1)^n$ theory by taking a combination of Chern-Simons and BF actions, the mixing being achieved with the help of a collection of integer coupling constants. Based on the Deligne-Beilinson cohomology, a partition function can then be computed for such a $\mathrm{U}(1)^n$ Chern-Simons theory. This partition function is clearly a topological invariant of the closed oriented $3$-manifold on which the theory is defined. Then, by applying a reciprocity formula a new expression of this invariant is obtained which should be a Reshetikhin-Turaev invariant. Finally, a duality between $\mathrm{U}(1)^n$ Chern-Simons theories is demonstrated. 
		\end{abstract}
				
	\end{frontmatter}
	
    \newpage
	
	\tableofcontents
	
	\newpage

\section*{Introduction}

In 1974, Shiing-Shen Chern and James Simons \cite{CS1974} introduced what is now known as the Chern-Simons $3$-form
\[
L\left[A\right] 
= \frac{1}{8\pi^{2}}\mathrm{Tr}\left(A\wedge F_{A} - \frac{1}{3}A\wedge A\wedge A\right)
= \frac{1}{8\pi^{2}}\mathrm{Tr}\left(A\wedge dA + \frac{2}{3}A\wedge A\wedge A\right)
\]
in their study of secondary characteristic classes. In this formula, $A$ is a $G$-connection (let us take $G = \mathrm{SU}(2) $ for further convenience) over a $3$-manifold $M$, and $F_{A} = dA + A\wedge A$ is its associated curvature $2$-form. The Chern-Simons $3$-form $L\left[A\right]$ can be viewed as an antiderivative of the second Chern class, i.e.,
\[
dL\left[A\right] 
= \frac{1}{8\pi^{2}}\mathrm{Tr}\left(F_{A}\wedge F_{A}\right).
\]
If we interpret the $\mathrm{SU}(2)$-connection $A$ as a physical field, the Chern-Simons $3$-form can be regarded as the Lagrangian of a physical system, and integrating it over $M$ defines action functionals 
\[
S^{\mathrm{SU}(2)}_{\mathrm{CS}_{k}}\left[A\right]
= k\int_{M}L\left[A\right] 
= \frac{k}{8\pi^{2}}\int_{M}\mathrm{Tr}\left(A\wedge dA + \frac{2}{3}A\wedge A\wedge A\right)
\]
called $\mathrm{SU}(2)$ Chern-Simons actions, which exhibit remarkable properties. First, a gauge transformation 
\[
A\longmapsto A^{g} = g^{-1}Ag + g^{-1}dg,\quad g\in C^{\infty}\left(M,\mathrm{SU}(2)\right)
\]
does not leave $S^{\mathrm{SU}(2)}_{\mathrm{CS}_{k}}$ invariant, but leaves it invariant up to a Wess-Zumino term
\[
S^{\mathrm{SU}(2)}_{\mathrm{WZ}_{k}}\left[g\right]
=\frac{k}{24\pi^{2}}\int_{M}\mathrm{Tr}\left(\left(g^{-1}dg\right)^{3}\right),
\]
which turns out to be an integer for a closed $3$-manifold $M$, provided the coupling constant $k$ is quantized, i.e., $k\in\Z$. In the formalism of path integral, rather than $S^{\mathrm{SU}(2)}_{\mathrm{CS}_{k}}$, we are interested in $e^{2\pi i S^{\mathrm{SU}(2)}_{\mathrm{CS}_{k}}}$, so that this object is gauge invariant. In this sense, the $\mathrm{SU}(2)$ Chern-Simons action is not a usual classical action, but rather some sort of ``purely quantum action''. Second, the Chern-Simons action does not contain any metric. In this sense, it is topological.

In 1978, Albert Schwartz \cite{Sch1978} showed how the Reidemeister/Ray-Singer/analytic torsion of a manifold, which is an important topological invariant independent of the choice of a metric, could arise from path integrals involving quadratic gauge invariant action functionals. This opened the way to the concept of Topological Quantum Field Theory (TQFT) in the formalism of the path integral.

In the 80s-90s, the works of Edward Witten \cite{Wit1982}, Simon Donaldson \cite{Don1983}, Michael Atiyah \cite{Ati1987} and Graeme Segal \cite{Seg2001} extended the idea that, with specific action functionals, path integrals could be interpreted as topological invariants. Since path integrals are generally ill-defined, the definition of Topological Quantum Field Theory through the path integral was abandoned to the advantage of the so-called Atiyah-Segal axioms which are expressed in the formalism of Category Theory, preserving the ``cutting and gluing formula'', a key idea in the formalism of path integral. More precisely, a Topological Quantum Field Theory is defined as a monoidal functor $\mathcal{Z}$ from the monoidal category $\left(\textsf{Bord}_{n},\sqcup\right)$, whose objects are closed $\left(n-1\right)$-manifolds (endowed with disjoint union) and morphisms are $n$-dimensional bordisms between them, to the category $\left(\textsf{Vect}_{\Cplx},\otimes\right)$, whose objects are complex vector spaces (endowed with the tensor product) and morphisms are linear maps between them. The vector spaces in the target can be interpreted as Hilbert spaces that occur in QFT in the Hamiltonian formalism.

Note that a closed $n$-manifold $M$ can be regarded as a bordism between two empty closed $\left(n-1\right)$-manifolds $\varnothing$, and Atiyah-Segal axioms impose that $\mathcal{Z}(\varnothing) = \Cplx$. As a consequence, the topological invariant $\mathcal{Z}(M)$ is a number, called the partition function of $M$.

In 1989, Edward Witten \cite{Wit1989} claimed that the partition function 
\[
\mathcal{Z}^{\mathrm{SU}(2)}_{\mathrm{CS}_{k}}(M) 
=\int_{\mathrm{Conn}_{\mathrm{SU}(2)}(M)}
\hspace{-0.40cm}\mathscr{D}\!A\,
e^{2\pi iS^{\mathrm{SU}(2)}_{\mathrm{CS}_{k}}\left[A\right]},
\]
of an $\mathrm{SU}(2)$ Chern-Simons theory for a coupling constant $k\in\Z$ over a closed $3$-manifold $M$ could actually be interpreted as the Jones polynomial $P_{K} = P_{K}(q)$ of a surgery knot $K$ of $M$ in the variable $q = e^{\frac{i\pi}{k+2}}$. 

Equivalently, $\mathcal{Z}^{\mathrm{SU}(2)}_{\mathrm{CS}_{k}}(M)$ can be interpreted as the Reshetikhin-Turaev invariant $\mathcal{Z}^{\mathcal{U}_{q}(\mathfrak{sl}_{2}(\Cplx))}_{\mathrm{RT}}(M)$ built from the modular category of representations of $\mathcal{U}_{q}(\mathfrak{sl}_{2}(\Cplx))$, the quantum deformation of the universal enveloping algebra of $\mathfrak{sl}_{2}(\Cplx)$ at the root of unity $q = e^{\frac{i\pi}{k+2}}$. This invariant is obtained from a knot diagram of an integral surgery knot of $M$ by labelling the strands with representations of $\mathcal{U}_{q}(\mathfrak{sl}_{2}(\Cplx))$. It can be derived from a Topological Quantum Field Theory in the sense of the Atiyah-Segal axioms, which we will call ``Reshetikhin-Turaev theory''.

The $\mathrm{SU}(2)$ Chern-Simons theory is an interesting example of nonabelian TQFT, but we can wonder what happens in the \textit{a priori} simpler abelian case. However, several games are possible. Indeed, $\mathrm{SU}(2)$ is compact and simply connected, and there is no abelian group satisfying these two assumptions simultaneously. The real line $\R$ is simply connected but not compact, while $\mathrm{U}(1)$ is compact but not simply connected. With the $\R$-principal bundles being trivializable, the usual gauge fixing procedure works the same as in the $\mathrm{SU}(2)$ case. Regarding the $\mathrm{U}(1)$ case, if we want to retain the idea of QFT that consists of summing over \textit{all} the possible configurations, then we have to deal with all the non-trivial $\mathrm{U}(1)$-principal bundles. This case was discussed in different and complementary manners in \cite{Man1998a,Man1998b,GMMS2005,BM2005,BGST2005,GT2008,GT2013,GT2014,MT2016JMP,MT2016NPB}. The natural extension to $\mathrm{U}(1)^{n}$ is discussed in \cite{Man1998a,Man1998b,GMMS2005,BM2005} and we aim to provide a more constructive complementary approach by generalizing the works of \cite{BGST2005,GT2008,GT2013,GT2014,MT2016JMP,MT2016NPB}.

In the first section, we will recall important facts on the topology of closed $3$-manifolds and Dehn surgery. In the second section, after recalling the relevant (to our case) facts on Deligne-Beilinson cohomology, we will derive the partition function of the $\mathrm{U}(1)^{n}$ Chern-Simons theory. In particular, we will see how the $\mathrm{U}(1)$ BF theory introduced in earlier papers turns out to be a specific case of $\mathrm{U}(1)^{2}$ Chern-Simons theory. We will also see that $\mathrm{U}(1)^{n}$ Chern-Simons theory exhibits a mixing of $\mathrm{U}(1)$ Chern-Simons and BF theories. Such a mixing is of interest in condensed matter physics. A reciprocity formula that connects the partition function directly with the surgery data will be presented. Based on this reciprocity formula, a duality will be eventually highlighted.

\textbf{Results}: In this article, we determine the partition function of the $\mathrm{U}(1)^{n}$ Chern-Simons Theory, as first discussed in \cite{TM2023}, where fields are Deligne-Beilinson cohomology classes. This partition is an invariant of the oriented closed $3$-manifold on which the theory is defined. Finally, by applying the Deloup-Turaev reciprocity formula \cite{Deloup2007} to this partition function we reveal a duality between $\mathrm{U}(1)^{n}$ Chern-Simons Theories.

\section{Facts on the topology of closed oriented $3$-manifolds}

Let us recall that any closed oriented $3$-manifold $M$ can be obtained by an integral surgery of $S^{3}$ along a framed link $\boldsymbol{\mathcal{L}}\subset S^{3}$ \cite{Lic1962,Wal1960}. Such a link has $n$ different components, $\mathcal{K}_{i}$, which are non intersecting oriented knots in $S^{3}$, a knot being an embedding of the circle $S^1$ into $S^3$. We write $\boldsymbol{\mathcal{L}} = \mathcal{K}_{1}\sqcup\hdots\sqcup \mathcal{K}_{n}$. The framing of the link assigns to each of its components an integer which is referred to as the charge of the component.

\begin{figure}
     \centering
\begin{subfigure}[b]{0.4\textwidth}
         \centering         
         \includegraphics[width=\textwidth]{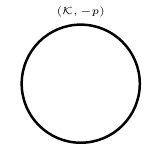}
         \caption{A framed unknot in $S^{3}$, with framing $-p$, along which the surgery of $S^{3}$ gives the lens space $L\left(p,1\right)$. For instance, $L\left(0,1\right) \cong S^{1}\times S^{2}$ and $L\left(2,1\right) \cong \RP^{3}$.}
         \label{Fig:Integral_Framed_Unknot}
     \end{subfigure}
     \hfill
\begin{subfigure}[b]{0.4\textwidth}
         \centering         
         \includegraphics[width=1.3\textwidth , trim=15 0cm -15 0]{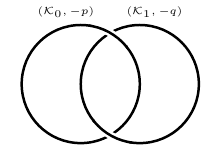}
         \caption{A framed Hopf link in $S^{3}$, with (coprime) framings $-p$ and $-q$ and linking number $1$, along which the surgery of $S^{3}$ gives the lens space $L\left(pq-1, q\right)$.}
         \label{Fig:Framed_Hopf_Link}
\end{subfigure}
     \hfill

\begin{subfigure}[b]{0.6\textwidth}
         \centering         
         \includegraphics[width=1\textwidth]{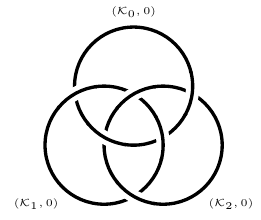}
         \caption{A framed Borromean link in $S^{3}$, whose components all carry a framing $0$, along which the surgery of $S^{3}$ gives the $3$-torus $T^{3} = S^{1}\times S^{1}\times S^{1}$. For each pair of components of the Borromean link the linking number is $0$, however, due to the way the components are connected, there is no ambient isotopy between the Borromean link and three unlinked unknots.}
         \label{Fig:Framed_Borromean_Link}
\end{subfigure}
\caption{Surgery presentation of various manifolds.}
\end{figure}

A framed link generates a collection of integers made of the linking numbers $\enlacement{\mathcal{K}_{i}}{\mathcal{K}_{j}}$ of its components and of the charges $e_i$. In fact, the charge $e_i$ of the component $\mathcal{K}_{i}$ can be seen as the self-linking number of $\mathcal{K}_i$: $\enlacement{\mathcal{K}_{i}}{\mathcal{K}_{i}}=e_i$. Figures
 \ref{Fig:Integral_Framed_Unknot}-\ref{Fig:Framed_Borromean_Link} present three different examples of integral surgery links in $S^{3}$. In the first example, the link on which the surgery is performed has only one component and is usually referred to as the unknot. In the second example the link is the Hopf link and in the last example it is the Borromean link.


\begin{figure}

\end{figure}



\subsection{The linking matrix of a surgery link}
Let $\mathcal{M}_{n}(\Z)$ denote the space of $n\times n$ integral matrices and $\mathrm{Sym}_{n}(\Z)$ the subset of those matrices that are symmetric.
\begin{defi}
\label{Def:Linking_Matrix}
Let $\boldsymbol{\mathcal{L}} = \mathcal{K}_{1}\sqcup\hdots\sqcup\mathcal{K}_{n}$ be an oriented framed link in $S^{3}$, the $i$-th component being
framed by $e_{i}\in\Z$. The integral $n\times n$ matrix $\mathbf{L}_{\boldsymbol{\mathcal{L}}} = \left(L_{ij}\right)_{1\leq i,j\leq n}\in\mathcal{M}_{n}(\Z)$, whose entries are
\[
L_{ij}
= \begin{cases} 
      e_{i} & \mbox{ if } i=j\\
      \enlacement{\mathcal{K}_{i}}{\mathcal{K}_{j}} & \mbox{ if } i\neq j\\ 
\end{cases}
\]
is called the linking matrix of $\boldsymbol{\mathcal{L}}$. More precisely, since $L_{ij} = \enlacement{\mathcal{K}_{i}}{\mathcal{K}_{j}} = \enlacement{\mathcal{K}_{j}}{\mathcal{K}_{i}} = L_{ji}$, $\mathbf{L}_{\boldsymbol{\mathcal{L}}}\in\mathrm{Sym}_{n}(\Z)$.    
\end{defi}

As an example, let us consider the links represented on Figures \ref{Fig:Integral_Framed_Unknot}-\ref{Fig:Framed_Borromean_Link}. Their linking matrices are respectively
\[
\mathbf{L}_\text{a} =
\begin{pmatrix}
    -p
\end{pmatrix} \, , \, 
\mathbf{L}_\text{b} =
\begin{pmatrix}
    -p & 1  \\
    1 & -q 
\end{pmatrix} \, , \,
\mathbf{L}_\text{c} =
\begin{pmatrix}
    0 & 0 & 0 \\
    0 & 0 & 0 \\
    0 & 0 & 0
\end{pmatrix}
\]

The linking matrix $\mathbf{L}_{\boldsymbol{\mathcal{L}}}$ of a link $\boldsymbol{\mathcal{L}}$ does not contain all the information on the link $\boldsymbol{\mathcal{L}}$, even if its components are unknots. Indeed, $\mathbf{L}_{\boldsymbol{\mathcal{L}}}$ contains only the information of pairwise linkings. Think about the linking matrix of the Borromean rings. Its non-diagonal entries are all zero, but the three components all together are not isolated. Hence, the linking matrix will be the same as the linking matrix of three isolated unknots, whereas this trivial link is definitely not ambient isotopic to the Borromean rings.

There exists a set of moves $\{\mu_1,\mu_{2}\} $, called Kirby moves, which can be applied to a link such that the resulting manifold remains the same after integral surgery.
It has been proven that closed oriented $3$-manifolds obtained by integral surgery on framed links $\boldsymbol{\mathcal{L}}$ and $\boldsymbol{\mathcal{L}}'$ are homeomorphic by an orientation preserving homeomorphism if and only if $\boldsymbol{\mathcal{L}}'$ can be obtained from $\boldsymbol{\mathcal{L}}$ by a sequence of Kirby moves \cite{Kir1978}.

\begin{prop} [\cite{Sav2012}]
The effect of the Kirby moves on the linking matrix $\mathbf{L}_{\boldsymbol{\mathcal{L}}}$ is as follows. The move $\mu_{1}$ replaces $\mathbf{L}_{\boldsymbol{\mathcal{L}}}$ by
\[
\begin{pmatrix}
 & & & 0 \\
 & \mathbf{L}_{\boldsymbol{\mathcal{L}}} & & \vdots \\
 & & & 0 \\
0 & \hdots & 0 & \pm 1
\end{pmatrix}
\]
i.e. if $\boldsymbol{\mathcal{L}}' = \mu_{1}\left(\boldsymbol{\mathcal{L}}\right)$, denote $\mathbf{L}_{\boldsymbol{\mathcal{L}}} = \left(L_{ij}\right)_{1\leq i,j\leq n}$ and $\mathbf{L}_{\boldsymbol{\mathcal{L}}'} = \left(L'_{ij}\right)_{1\leq i,j\leq n+1}$, then 
\[
L'_{ij}
= \begin{cases} 
      L_{ij} & \mbox{ if } 1\leq i,j\leq n\\
      \pm 1 & \mbox{ if } i = j = n+1 \\
      0 & \mbox{ otherwise.}
\end{cases}
\]

The move $\mu_{2}$ slides $\mathcal{K}_{i_{0}}$ over $\mathcal{K}_{j_{0}}$ to produce the pair $\left(\mathcal{K}_{i_{0}}+\mathcal{K}_{j_{0}}\right)\sqcup\mathcal{K}_{j_{0}}$. The framing of the new component then becomes $\left(L_{i_{0}i_{0}} + L_{j_{0}j_{0}} \pm 2L_{i_{0}j_{0}}\right)$. This procedure is described more thoroughly in \cite{Sav2012}. The new linking matrix is obtained from $\mathbf{L}_{\boldsymbol{\mathcal{L}}}$ by adding (or subtracting) the $j_{0}$-th row to (from) the $i_{0}$-th row and the $j_{0}$-th column to (from) the $i_{0}$-th column, i.e., if $\boldsymbol{\mathcal{L}}' = \mu_{2}\left(\boldsymbol{\mathcal{L}}\right)$, $\mathbf{L}_{\boldsymbol{\mathcal{L}}} = \left(L_{ij}\right)_{1\leq i,j\leq n}$ and $\mathbf{L}_{\boldsymbol{\mathcal{L}}'} = \left(L'_{ij}\right)_{1\leq i,j\leq n}$, then
\[
L'_{ij}
= \begin{cases} 
      L_{ii_{0}}\pm L_{ij_{0}} & \mbox{ if } i\neq i_{0}, j=i_{0}\\
      L_{i_{0}j}\pm L_{j_{0}j} & \mbox{ if } i=i_{0}, j\neq i_{0}\\
      \left(L_{i_{0}i_{0}} + L_{j_{0}j_{0}} \pm 2L_{i_{0}j_{0}}\right)  & \mbox{ if } i=j=i_{0}\\
      L_{ij} & \mbox{ otherwise.}
\end{cases}
\]
The matrix $\mathbf{P}= \left(P_{ij}\right)_{1\leq i,j\leq n}$ such that $\mathbf{L}_{\boldsymbol{\mathcal{L}}'} = {}^{t}\mathbf{P}\mathbf{L}_{\boldsymbol{\mathcal{L}}}\mathbf{P}$ is
\[
\mathbf{P} = 
     \bordermatrix{ & P_{*1} & \cdots & P_{*i_{0}} & \cdots & P_{*j_{0}} & \cdots & P_{*n} \cr
       P_{1*} & 1  & \cdots & 0 & \cdots & 0 &\cdots & 0  \cr
       \vdots & \vdots &  \ddots & \vdots &\ddots & \vdots  &\ddots & \vdots  \cr
       P_{i_{0}*} & 0 & \cdots & 1 & \cdots & 0 &\cdots & 0 \cr
       \vdots & \vdots &  \ddots & \vdots &\ddots & \vdots  &\ddots & \vdots  \cr
       P_{j_{0}*} &  0 & \cdots & \pm 1 & \cdots & 1 & \cdots & 0  \cr
       \vdots & \vdots &  \ddots & \vdots &\ddots & \vdots  &\ddots & \vdots \cr
       P_{n*} & 0 & \cdots & 0 & \cdots & 0 & \cdots & 1 } \qquad
\]
\end{prop}

\begin{prop}
\label{canoformL}
(\cite{Mou2015})
Let $\boldsymbol{\mathcal{L}}$ be an integral surgery link of a closed oriented $3$-manifold $M$. Then there exists a sequence of Kirby moves producing a new surgery link $\boldsymbol{\mathcal{L}}'$ such that
\[
\mathbf{L}_{\boldsymbol{\mathcal{L}}'}
=\begin{pmatrix*}[l]
& \mathbf{L}_{0} & &&  \boldsymbol{0}_{r,b_{1}} & \\
 \\
& \boldsymbol{0}_{b_{1},r} & &&  \boldsymbol{0}_{b_{1}} & \\
\end{pmatrix*}\in\mathrm{Sym}_{n}(\Z)
\]
where $\mathbf{L}_{0}\in\mathrm{Sym}_{r}(\Z)$ has nonzero determinant (the choice of index $b_{1}$ will be explained later on.)  An explicit proof can also be found on \cite{TM2023}.
\end{prop}

\begin{prop}
Any manifold obtained from surgery along a link $\boldsymbol{\mathcal{L}}$ can also be obtained by surgery along a link $\boldsymbol{\mathcal{L}}'$ whose components are all unknots. This result is mentioned explicitly in \cite{Yu2008} and can be deduced from the construction presented in \cite{Lic1962}. 
\end{prop}

\begin{prop} (\cite{Sav2012})
Any closed oriented $3$-manifold can be obtained by means of an even surgery in $S^3$, i.e., a surgery along a link whose framings are all even integers. The linking matrix of an even link is obviously even, i.e., a matrix whose diagonal elements are all even. A process showing how this can be done on the linking matrix, using only Kirby moves, can be found in the Appendix.
\end{prop}

\subsection{Homology and linking form from the linking matrix} 

Let us consider a closed oriented $3$-manifold obtained by integral surgery along a link $\boldsymbol{\mathcal{L}}\in S^{3}$ whose associated linking matrix $\mathbf{L}_{\boldsymbol{\mathcal{L}}}$ can be written as
\[
\mathbf{L}_{\boldsymbol{\mathcal{L}}}
=\begin{pmatrix}
\mathbf{L}_{0} & \boldsymbol{0} \\
\boldsymbol{0} & \boldsymbol{0} \\
\end{pmatrix}\in\mathrm{Sym}_{n}(\Z)
\]
where $\mathbf{L}_{0}\in\mathrm{Sym}_{r}(\Z)$ has nonzero determinant. Then $b_{1} = n-r$, i.e.,
\[
FH_{1}(M) 
\cong\Z^{n-r}
\]
and 
\[
TH_{1}(M) 
\cong \slfrac{\Z^{r}}{\mathbf{L}_{0}\Z^{r}},
\]
so that
\[
H_{1}(M) 
\cong\slfrac{\Z^{n}}{\mathbf{L}\Z^{n}}
\cong\Z^{n-r}\oplus\slfrac{\Z^{r}}{\mathbf{L}_{0}\Z^{r}}.
\]

In the following, we will consider the invariant factors decomposition, i.e., when writing $TH_{1}(M)\cong\slfrac{\Z^{r}}{\mathbf{L}_{0}\Z^{r}}\cong\Z_{p_{1}}\oplus\hdots\oplus\Z_{p_{t}}$, for some $t \leq r$ and $p_{i}\vert p_{i+1}$ \footnote{An explicit algorithm for this procedure can be found in \cite{TM2023}}.
\begin{prop}
For $M$ a closed oriented $3$-manifold, by Poincaré duality and the universal coefficient theorem \cite{BT1982}, $b_{0} = b_{3}$, $b_{1} = b_{2}$, $TH_{2}(M)\cong TH^{1}(M)\cong 0$, and $H_{1}(M)\cong H^{2}(M)$, i.e.,
\begin{itemize}
    \item[-] $H_{0}(M)\cong H_{3}(M)\cong H^{0}(M)\cong H^{3}(M)\cong\Z$
    \item[-] $H_{1}(M)\cong H^{2}(M)\cong\Z^{b_{1}}\oplus\Z_{p_{1}}\oplus\hdots\oplus\Z_{p_{t}}$
    \item[-] $H_{2}(M)\cong H^{1}(M)\cong\Z^{b_{1}}$
\end{itemize}    
\end{prop}

As we saw above, if we determine $H_{1}(M)$ then we know all the homology and the cohomology of $M$. Of course, this is not sufficient to classify $M$ up to homeomorphism or homotopy equivalence. The fundamental group $\pi_{1}(M)$ is a much stronger invariant (whose abelianization is $H_{1}(M)$), but it turns out not to be sufficient either. Indeed, it cannot classify the lens spaces. The following invariant can make this distinction: 
\begin{defi} [\cite{GL1978}]
The linking form of $M$ is defined homologically as
\[
\begin{tabular}{rccl}
   $\mathbf{Q}:$ 
   & $TH_{1}(M)\times TH_{1}(M)$ & $\longrightarrow$ & $ \slfrac{\Q}{\Z}$ \\
   & & & \\
   & $\left(\left[\gamma_{1}\right],\left[\gamma_{2}\right]\right)$ 
   & $\longmapsto$ & $\left(\frac{\gamma_{1}\pitchfork \Sigma_{2}}{p}\right)^{\#} =\mathbf{Q}(\left[\gamma_{1}\right],\left[\gamma_{2}\right])$ 
\end{tabular}
\]
where $\Sigma_{2}$ is an integral $2$-chain such that $p\gamma_{2} = \partial\Sigma_{2}$, $\gamma_{1}$ and $\gamma_{2}$ being representatives of $\left[\gamma_{1}\right]$ and $\left[\gamma_{2}\right]$ respectively, and where $\pitchfork$ denotes the transverse intersection and $(\;)^{\#}$ the intersection number.

By duality, it can be defined cohomologically as
\[
\begin{tabular}{rccl}
$\mathbf{Q}:$
  & $TH^{2}(M)\times TH^{2}(M)$ & $\longrightarrow$ & $ \slfrac{\Q}{\Z}$ \\
  & & & \\
  & $\left(\left[a_{1}\right], \left[a_{2}\right]\right)$ 
  & $\longmapsto$ & $\left(\frac{C_{1}\smile a_{2}}{p}\right) (M) = \mathbf{Q}(\left[a_{1}\right], \left[a_{2}\right])$ 
\end{tabular}
\]
where $C_{1}$ is an integral $1$-cochain such that $dC_{1} = pa_{1}$, $a_{1}$ and $a_{2}$ being representatives of $\left[a_{1}\right]$ and $\left[a_{2}\right]$ respectively, $\smile$ denotes the cup product and the whole expression being evaluated over the $3$-manifold $M$ (seen as a $3$-cycle).

\end{defi}

\begin{rema}
We have decided to denote both the homological and cohomological linking forms with the same letter. The context will clearly indicates which form is being considered. In the cohomological description, $dC_{1} = pa_{1}$ means that $\frac{C_{1}}{p}$ is a $1$-cocycle with coefficients in $\slfrac{\Q}{\Z}$. By the Leibniz rule for the differential $d$, we have:
\[
d\left(C_{1}\smile a_{2}\right) = dC_{1}\smile a_{2} - C_{1}\smile da_{2} = pa_{1}\smile a_{2} 
\]
since $a_{2}$ is an integral $2$-cocycle and $dC_{1} = pa_{1}$. Hence, $C_{1}\smile a_{2}$ is an integral $3$-cochain, but $\frac{C_{1}\smile a_{2}}{p}$ is a $3$-cocycle with coefficients in $\slfrac{\Q}{\Z}$, so that its cohomology class is just an element of $\slfrac{\Q}{\Z}$. Naturally, the duality mentioned in the above definition is a Poincaré duality.  
\end{rema}

\begin{prop}[\cite{Sav2012},\cite{Mou2015},\cite{GL1978},\cite{Kyl1954},\cite{Kyl1959} ]
Assume a closed oriented $3$-manifold is obtained from $S^{3}$ by integral surgery along a link $\boldsymbol{\mathcal{L}}\in S^{3}$ such that 
\[
\mathbf{L}_{\boldsymbol{\mathcal{L}}}
=\begin{pmatrix}
\mathbf{L}_{0} & \boldsymbol{0} \\
\boldsymbol{0} & \boldsymbol{0} \\
\end{pmatrix}\in\mathrm{Sym}_{n}(\Z)
\]
where $\mathbf{L}_{0}\in\mathrm{Sym}_{r}(\Z)$ has nonzero determinant. Then
\[
\mathbf{Q} = \mathbf{L}^{-1}_{0}
\]
\end{prop}

\begin{rema}
Two linking matrices $\mathbf{L}_{1}$ and $\mathbf{L}_{2}$ related by Kirby moves will produce two different matrices $\mathbf{Q}_{1}$ and $\mathbf{Q}_{2}$, but these matrices will evaluate in the same manner modulo $\Z$. This is especially easy to see with first Kirby moves, which add or delete generators of trivial homology. The block associated with those moves is just a diagonal of $\pm 1$ which remains the same under inversion. Evaluating this block on some generators of trivial homology (essentially the group $\Z/\Z \cong \{0\}$) returns an integer which disappears modulo $\Z$. The same happens with the second Kirby move, although it is less easy to see.
\end{rema}

\begin{rema}
We can have two closed oriented $3$-manifolds with the same $\mathbf{Q}$ but completely different free homology, as this homology sector is not seen by $\mathbf{Q}$.
\end{rema}

\begin{rema}
In the next section, we will see that the partition function is built on $\mathbf{Q}$, not $\mathbf{L}$, and that $\mathbf{L}$ enters only through a reciprocity formula. The two objects, have an important distinction.
Consider a closed oriented $3$-manifold with no free first homology whose linking matrix is $\mathbf{L}$. Since $\mathbf{L}$ is an integral $m\times m$ symmetric matrix (for some $m$), its inverse is a rational ($m\times m$) symmetric matrix which defines a representative $\mathbf{Q}$ 
 of the linking form on $TH_1(M) \simeq \Z^m /\mathbf{L} \Z^m$. However, conversely, if we know a representative $\mathbf{Q}$ of the linking form, it is not straightforward to map it to an integral matrix $\mathbf{L}$ such that $\mathbf{L}$ is the linking matrix of the manifold. The crucial difference is that if we take an arbitrary representative $\mathbf{Q}$ of the linking form, its inverse will generally not be an integral matrix. So the process of deriving the linking matrix $\mathbf{L}$ of an integral surgery link from $\mathbf{Q}$ is more involved than just inverting $\mathbf{Q}$. Note that as previously mentioned, neither $\mathbf{L}$ nor $\mathbf{Q}$ are unique, even their dimensions can vary and most notably, we can have cases where the smallest dimension for $\mathbf{Q}$ can be smaller than the smallest dimension for $\mathbf{L}$ as in the example below.


A small example is the lens space $L\left(5,2\right)$, the smallest (dimensionally) integral surgery producing this space has linking matrix $\mathbf{L} = \begin{pmatrix}
    3 & 1\\
    1 & 2
\end{pmatrix}$ while its "simplest" linking form is $\mathbf{Q} = \frac{2}{5}$ acting on $\Z/5\Z$. There is no way to invert $\mathbf{Q}$ here to get a $1$-dimensional integral matrix. However, $\mathbf{L}^{-1}$ acting on $\mathbb{Z}^2/\mathbf{L}\mathbb{Z}^2$ is also a linking form and its inverse is clearly an integral symmetric matrix. Note that $\Z/5\Z \simeq \mathbb{Z}^2/\mathbf{L}\mathbb{Z}^2 \simeq TH_1\left(L\left(5,2\right)\right) $.
\end{rema}

\section{$\mathrm{U}(1)^{n}$ Chern-Simons theory}

Contrary to $\mathrm{SU}(2)$, the gauge group $\mathrm{U}(1)$ is not simply connected. Hence, there are in general nontrivializable $\mathrm{U}(1)$-bundles, and the $\mathrm{U}(1)$ connections cannot, in general, be written as (global) $1$-forms (with coefficients in $\mathrm{Lie}\left(\mathrm{U}(1)\right) = i\R$) over the base closed oriented $3$-manifold $M$. Moreover, contrary to the standard approach of quantum field theory involving gauge fixing, we want to work here directly with the fields modulo gauge transformations, i.e., gauge classes of $\mathrm{U}(1)$-connections. Deligne-Beilinson cohomology is an appropriate mathematical framework for describing  $\mathrm{U}(1)$ gauge classes without referring to anything other than $M$.

\subsection{Deligne-Beilinson cohomology}

As the construction of the Deligne-Beilinson cohomology groups is quite irrelevant for us, we will only mention results that are important in expressing the $\mathrm{U}(1)^n$ Chern-Simons action as well as in determining the corresponding partition function.

\begin{figure}
\begin{center}
\includegraphics[scale = 1.]{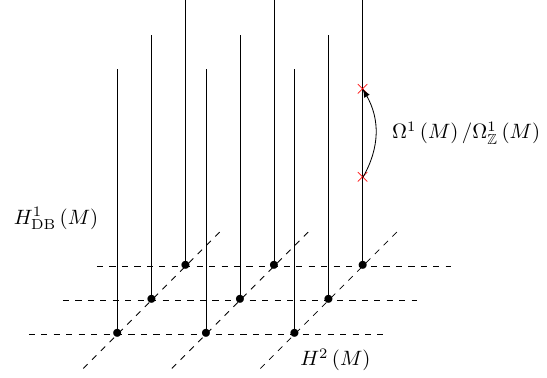}
\caption{Representation of $H^{1}_{\mathrm{DB}}(M)$}
\label{Space_Fields}
\end{center}
\end{figure}

\begin{prop}[\cite{CS1985},\cite{B1993},\cite{HLZ03},\cite{HS2005},\cite{SS2008} ]
The space of gauge classes of $\mathrm{U}(1)$ connections is the first group of Deligne-Beilinson cohomology $H^{1}_{\mathrm{DB}}(M)$. It is a $\Z$-module which sits in the following short exact sequence that splits:
\begin{equation}
\label{SES_1}
0
\to \slfrac{\Omega^{1}(M)}{\Omega^{1}_{\Z}(M)}
\to H^{1}_{\mathrm{DB}}(M)
\to H^{2}(M)
\to 0
\end{equation}
In this short exact sequence, $\Omega^{1}(M)$ is the space of $1$-forms over $M$, $\Omega^{1}_{\Z}(M)$ is the space of closed $1$-forms with integral periods over $M$, and $H^{2}(M)$ is the second cohomology group of $M$. Figure \ref{Space_Fields} is a way to visualize this exact sequence.
\end{prop}

Let us briefly explain the above exact sequence. Any $\mathrm{U}(1)$ gauge field of $M$ is actually a well defined $1$-form on some $\mathrm{U}(1)$ principal bundle over $M$. This way, to any $\mathrm{U}(1)$ gauge class corresponds a class of isomorphic $\mathrm{U}(1)$ principal bundles over $M$. Now, the set classes of isomorphic $\mathrm{U}(1)$ principal bundles over $M$ is canonically identified with $H^{2}(M)$, the second cohomology group of $M$. This implies that to each DB class we can associate a class in $H^{2}(M)$, which describes the second non-trivial group homomorphism of \eqref{SES_1}. Furthermore, any $1$-form $\omega$ of $M$ trivially defines a $\mathrm{U}(1)$ gauge field. Yet, if $\omega$ is closed with integral periods, then there exist a $\mathrm{U}(1)$-valued function $g$ on $M$ such that $\omega = g^{-1} d g$. In such a case the gauge field is a gauge transformation and as such its gauge class is zero. The quotient $\slfrac{\Omega^{1}(M)}{\Omega^{1}_{\Z}(M)}$ thus trivially maps into $H^{1}_{\mathrm{DB}}(M)$, which yields the first non-trivial homomorphism of \eqref{SES_1}. The exactness of this sequence is mainly ensured firstly by the fact that the de Rham cohomology class of the curvature of a $\mathrm{U}(1)$ gauge field defines an element of $FH^{2}(M)$, the free sector of $H^{2}(M)$, and secondly by the fact that the curvature of a $1$-form being the de Rham derivative of this $1$-form, its de Rham cohomology class is zero.

Due to the possible presence of a torsion sector in $H^{2}(M)$, the fact that sequence \eqref{SES_1} splits is not completely trivial. Let us give a sketch of proof of this splitting. For each $\mathbf{n} \in H^{2}(M)$ we consider the ``fiber'' made of all the classes in $H^{1}_{\mathrm{DB}}(M)$ whose associated cohomology class is $\mathbf{n}$. For each $\mathbf{n} \in H^{2}(M)$ we select a DB class, $A_{\mathbf{n}}$, on the fiber over $\mathbf{n}$, the other elements of this fiber being reached by adding to $A_{\mathbf{n}}$ the elements of $\slfrac{\Omega^{1}(M)}{\Omega^{1}_{\Z}(M)}$ (or rather their trivial images in $H^{1}_{\mathrm{DB}}(M)$). On the fiber over the trivial cohomology class, usually referred to as the trivial fiber, the zero DB class plays the role of a canonical origin, the trivial fiber being thus canonically identified with $\slfrac{\Omega^{1}(M)}{\Omega^{1}_{\Z}(M)}$. More surprisingly, on the torsion fibers, i.e., the fibers over the elements of $TH^2(M)$, there exist ``pseudo-canonical'' origins which were introduced in \cite{GT2013}. These origins are DB classes $\mathbf{A}_{\boldsymbol{\tau}}$ such that if $p.\boldsymbol{\tau} = 0$ then $p.\mathbf{A}_{\boldsymbol{\tau}} = \mathbf{0}$. On fibers over $FH^{2}(M)$ there are no particular origins so we can pick as origin on these ``free'' fibers any DB class we want. It is not difficult to show that with such a choice of origins on the fibers over $H^{2}(M)$ we obtain a group homomorphism $\sigma:H^{2}(M) \to H^{1}_{\mathrm{DB}}(M)$ which, by construction, is the right inverse of the second non-trivial morphism of exact sequence \eqref{SES_1}.

\begin{cons}
We can write
\begin{align}
H^{1}_{\mathrm{DB}}(M)
\cong \slfrac{\Omega^{1}(M)}{\Omega^{1}_{\Z}(M)} \oplus H^{2}(M)
\end{align}
\end{cons}

\begin{prop}
The short sequence
\begin{equation}
0
\to \slfrac{\Omega^{1}_{\mathrm{cl}}(M)}{\Omega^{1}_{\Z}(M)}
\to \slfrac{\Omega^{1}(M)}{\Omega^{1}_{\Z}(M)}
\to \slfrac{\Omega^{1}(M)}{\Omega^{1}_{\mathrm{cl}}(M)}
\to 0
\end{equation}
where $\Omega^{1}_{\mathrm{cl}}(M)$ is the space of closed $1$-forms, is exact and splits. And so:
\[
\slfrac{\Omega^{1}(M)}{\Omega^{1}_{\Z}(M)} \cong \slfrac{\Omega^{1}_{\mathrm{cl}}(M)}{\Omega^{1}_{\Z}(M)} \oplus \slfrac{\Omega^{1}(M)}{\Omega^{1}_{\mathrm{cl}}(M)}
\]
\end{prop}

\begin{rema}
We can show that
\begin{equation}
\slfrac{\Omega^{1}_{\mathrm{cl}}(M)}{\Omega^{1}_{\Z}(M)} 
\cong\left(\slfrac{\R}{\Z}\right)^{b_{1}}
\end{equation}
while, by definition,
\begin{equation}
FH^{2}(M)
\cong\Z^{b_{1}}
\end{equation}
\end{rema}

\begin{cons}
We can finally write \cite{HLZ03}
\begin{align}
\label{decompH_DB}
H^{1}_{\mathrm{DB}}(M)
\cong \slfrac{\Omega^{1}(M)}{\Omega^{1}_{\mathrm{cl}}(M)}
\oplus\slfrac{\Omega^{1}_{\mathrm{cl}}(M)}{\Omega^{1}_{\Z}(M)} 
\oplus FH^{2}(M)
\oplus TH^{2}(M)
\end{align}
and thus we can decompose $A\in H^{1}_{\mathrm{DB}}(M)$ as 
\begin{align}
\label{decompA}
A = \underset{\underset{FH^{2}(M)}{\rotatebox[origin=c]{-90}{$\in$}}}{A_{\mathbf{m}}} 
+ \underset{\underset{TH^{2}(M)}{\rotatebox[origin=c]{-90}{$\in$}}}{A_{\boldsymbol{\kappa}}} 
+ \underset{\underset{\slfrac{\Omega^{1}(M)}{\Omega^{1}_{\mathrm{cl}}(M)}}
{\rotatebox[origin=c]{-90}{$\in$}}}{\alpha_{\perp}} 
+ \underset{\underset{\slfrac{\Omega^{1}_{\mathrm{cl}}(M)}{\Omega^{1}_{\Z}(M)}}
{\rotatebox[origin=c]{-90}{$\in$}}}{\alpha_{0}}
\end{align}
\end{cons}
As the above isomorphism is not canonical, this decomposition is not unique. Nevertheless, the computations we want to perform are independent from the choices that yield decomposition \eqref{decompA} \cite{GT2014}. 
The whole point is to find a decomposition that makes our computations easy. The pseudo-canonical origins on torsion fibers were precisely introduced with this objective of simplicity in mind. Before giving such a decomposition,  let us  introduce two important operations.

\subsection{Operations on Deligne-Beilinson cohomology}

\begin{prop}
Over $H^{1}_{\mathrm{DB}}(M)$, there exists a (symmetric) pairing
\begin{equation}
\begin{tabular}{rccc}
$\star:$ & $H^{1}_{\mathrm{DB}}(M)\times H^{1}_{\mathrm{DB}}(M)$
& $\longrightarrow$ 
& $\slfrac{\Omega^{3}(M)}{\Omega^{3}_{\Z}(M)}$ \\
& \\
& $\left(A,B\right)$ 
& $\longmapsto$
& $A\star B$
\end{tabular}
\end{equation}
\end{prop}

\begin{prop}
Another notion of integral will be very useful here. This is
\begin{equation}
\label{Int_M3}
\begin{tabular}{rccc}
$\int_{M}:$
& $\slfrac{\Omega^{3}(M)}{\Omega^{3}_{\Z}(M)}$
& $\longrightarrow$ 
& $\slfrac{\R}{\Z}$ \\
& \\
& $L$ 
& $\longmapsto$
& $\int_{M}L$
\end{tabular}
\end{equation}
\end{prop}

\begin{rema}
Let us emphasize that the integral \eqref{Int_M3} is computed with the help of representatives of the source space. Due to the quotient structure of this source space, this integral is ill-defined in $\R$ whereas it is well defined as an $\slfrac{\R}{\Z}$-valued functional. In fact, in the partition function that we will introduce and study later in this article, it is the functional $e^{2 \pi i \int_{M}}$ that will appear, this $\mathrm{U}(1)$-valued functional hence being well-defined on $\slfrac{\Omega^{3}(M)}{\Omega^{3}_{\Z}(M)}$.
\end{rema}

The combination of the pairing $\star$ with decomposition (\ref{decompA}) yields the following crucial proposition.  

\begin{prop}[\cite{GT2008}, \cite{GT2013}, \cite{GT2014}, \cite{HMT2022_1}, \cite{HMT2022_2}] 
If we use the pseudo-canonical origins on the torsion fibers of $H^{1}_{\mathrm{DB}}(M)$, the components appearing in decomposition  \eqref{decompA} are such that
\begin{multicols}{2}
\begin{enumerate}[1)]
\item $\int_{M} \beta_{0} \star A_{\mathbf{m}_{A}} \underset{\Z}{=} \boldsymbol{\theta}_{B} \cdot \mathbf{m}_{A}$,
\item $\int_{M} \beta_{0}\star A_{\boldsymbol{\kappa}_{A}} \underset{\Z}{=} 0$,
\item $\int_{M} \beta_{0} \star \alpha_{0} \underset{\Z}{=} 0$,
\item $\int_{M} \beta_{0} \star \alpha_{\perp} \underset{\Z}{=} 0$,
\item $\int_{M}B_{\boldsymbol{\kappa}_{B}}\star A_{\boldsymbol{\kappa}_{A}}
\underset{\Z}{=} -\mathbf{Q}\left(\boldsymbol{\kappa}_{B},\boldsymbol{\kappa}_{A}\right)$,
\item $\int_{M}\beta_{\perp}\star A_{\boldsymbol{\kappa}_{A}} \underset{\Z}{=} 0$,
\end{enumerate}
\end{multicols}
\vspace{-0.25cm}
\noindent where $\boldsymbol{\theta}_{B} \in \left(\slfrac{\R}{\Z}\right)^{b_{1}}$, $\mathbf{m}_{A} \in \Z^{b_{1}}$, and $\mathbf{Q}:TH^{2}\times TH^{2}\to\slfrac{\Q}{\Z}$ is the linking form of $M$.
\end{prop}

Let us explain shortly the first of the above property. Any DB class $\beta_0$ can be obtained as follows. Let $(S_i)$ be a family of $2$-cycles that generate $H_2(M)$ and $(\gamma_i)$ a family of $1$-cycles that generate $FH_1(M)$  where the intersection number of $S_i$ with $\gamma_j$ is $\delta_{ij}$. To each $S_i$ we associate a closed $1$-form $\rho_i$ such that $\oint_{\gamma_j} \rho_i = \delta_{ij}$. Then, any $\beta_0$ is canonically identified with a combination $\sum_i \theta_i \rho_i$, where $\theta_i \in \slfrac{\R}{\Z}$. Now, if $A$ is a DB class then $\int_M \beta_0 \star A = \sum_i \theta_i \int_M \rho_i \wedge F(A) = \sum_i \theta_i \oint_{S_i} F(A)$, where in the last equality we used the property that $\rho_i$ is a Poincaré dual of $S_i$ and where $F(A)$ is the curvature of any gauge field whose class is $A$. Moreover, if $A$ belongs to a free fiber of $H^{1}_{\mathrm{DB}}(M)$, then its curvature represents an element of $FH^2(M)$ so that $\oint_{S_i} F(A) = m_i \in \Z$ for $i=1, \cdots , b_1$. Hence, $\int_M \beta_0 \star A = \sum_i \theta_i m_i = \vec{\theta} \cdot \vec{m}$ where $\vec{m}$ represents the class of $A$ in $FH^2(M) \cong \Z^{b_1}$.

\subsection{Generalization to gauge classes of $\mathrm{U}(1)^{n}$ connections}

From now on, we will consider column vectors
\begin{equation}
\mathbf{A} = 
\begin{pmatrix}
A_{1}\\
\vdots\\
A_{n}
\end{pmatrix}
\in\left(H^{1}_{\mathrm{DB}}(M)\right)^{n}
\end{equation} 
where $A_{i}\in H^{1}_{\mathrm{DB}}(M)$ for $i\in\left\lbrace 1,\hdots,n\right\rbrace$, which thus represent $\mathrm{U}(1)^{n}$-connections. Moreover, we extend the symmetric pairing $\star$ over $H^{1}_{\mathrm{DB}}(M)$ to a pairing over $\left(H^{1}_{\mathrm{DB}}(M)\right)^{n}$ by setting
\begin{equation}
\begin{tabular}{rccc}
$\star:$&$\left(H^{1}_{\mathrm{DB}}(M)\right)^{n}\times\left(H^{1}_{\mathrm{DB}}(M)\right)^{n}$
& $\longrightarrow$ 
& $\slfrac{\Omega^{3}(M)}{\Omega^{3}_{\Z}(M)}$ \\
& & \\
&$\left(\mathbf{A} = 
\begin{pmatrix}
A_{1}\\
\vdots\\
A_{n}
\end{pmatrix},
\mathbf{B} = 
\begin{pmatrix}
B_{1}\\
\vdots\\
B_{n}
\end{pmatrix}
\right)$ 
& $\longmapsto$
& ${}^{t}\mathbf{A}\star\mathbf{B} = \sum\limits_{i=1}^{n}A_{i}\star B_{i}$
\end{tabular}
\end{equation}
If $A\in H^{1}_{\mathrm{DB}}(M)$ and $C\in\Z$, then $CA \in H^{1}_{\mathrm{DB}}(M)$. More generally, if $\mathbf{A}\in\left(H^{1}_{\mathrm{DB}}(M)\right)^{n}$ and $\mathbf{C}\in\mathcal{M}_{n}(\Z)$, then $\mathbf{C}\mathbf{A}\in\left(H^{1}_{\mathrm{DB}}(M)\right)^{n}$, where the operation between $\mathbf{C}$ and $\mathbf{A}$ is understood as a product of matrices.

The $\mathrm{U}(1)^{n}$ Chern-Simons action is then defined as follows.

\begin{defi}\label{def_of_C_matrix}
We define the $\mathrm{U}(1)^{n}$ Chern-Simons action as the $\slfrac{\R}{\Z}$-valued functional
\begin{equation}
\begin{tabular}{rccl}
$S_{\mathrm{CS}_{\mathbf{C}}}:$ 
& $\left(H^{1}_{\mathrm{DB}}(M)\right)^{n}$ 
& $\longrightarrow$
& $\slfrac{\R}{\Z}$ \\
& \\
& $\mathbf{A}$ 
& $\longmapsto$
& $S_{\mathrm{CS}_{\mathbf{C}}}\!\left(\mathbf{A}\right) = \int_{M}{}^{t}\mathbf{A}\star\mathbf{C}\mathbf{A}$.
\end{tabular}
\end{equation}
where 
\begin{align*}
   \mathbf{C} = 
   \begin{pmatrix}
   k_{11} & k_{12} & \hdots & k_{1n} \\
   k_{21} & k_{22} & \hdots & k_{2n} \\
   \vdots & \vdots & \ddots & \vdots \\
   k_{n1} & k_{n2} & \hdots & k_{nn} \\
   \end{pmatrix}
\end{align*}
is an integral mixing-coupling matrix of the entries of $\mathbf{A}$. Note that the action is invariant under the transformation $( \mathbf{C} , M ) \rightarrow (- \mathbf{C} , -M ) $ where by $-M$ we denote the manifold with reverse orientation.
\end{defi}

\begin{rema}
The $\mathrm{U}(1)$ Chern-Simons theory defined by the one-dimensional matrix $\mathbf{C} = \left( k \right)$ is obviously the Chern-Simons theory studied in \cite{GT2014}. 
\end{rema}

\begin{rema}\label{BFinCS}
The $\mathrm{U}(1)^2$ Chern-Simons theory defined by the matrix
\[
\mathbf{C} = 
\begin{pmatrix}
0 & k_{12}\\
k_{21} & 0
\end{pmatrix}
\]
yields the $\mathrm{U}(1)$ BF theory with coupling constant $k= k_{12}+k_{21}$ studied in \cite{MT2016JMP}. Indeed, thanks to the commutativity of the $\star$-product, we can always turn the Chern-Simons action defined by the above matrix $\mathbf{C}$ into the Chern-Simons action defined by 
\[
\mathbf{C}' =
\begin{pmatrix}
0 & k\\
0 & 0
\end{pmatrix} 
\]
This property is reminiscent of how the $\Z_{k}$ Turaev-Viro invariant can be obtained as a Reshetikhin-Turaev invariant constructed on the Drinfeld center of $\Z_{k} \times \Z_{k} = \Z_{k}^2$.

More generally, the commutativity of the $\star$-product allows to turn the Chern-Simons action defined by any integral matrix $\mathbf{C}$ into the Chern-Simons action defined by an upper triangular integral matrix. This implies that there are $n(n+1)/2$ independent coupling constants in the $\mathrm{U}(1)^n$ Chern-Simons theory.
\end{rema}
\begin{rema}
    Even more generally, $\mathrm{U}(1)^n$ BF theory is a sub-case of $\mathrm{U}(1)^{2n}$ Chern-Simons theory \cite{TM2023}.
\end{rema}

\subsection{$\mathrm{U}(1)^{n}$ Chern-Simons partition function}

We want to study a functional integral of the form
\[
\mbox{``} \mathcal{Z}_{\mathrm{CS}_{\mathbf{C}}}
= \frac{1}{\mathcal{N}_{\mathrm{CS}_{\mathbf{C}}}(M)}
\int_{\left(H^1_{\mathrm{DB}}(M)\right)^{n}}\mathscr{D}\mathbf{A} 
e^{2\pi i S_{\mathrm{CS}_{\mathbf{C}}}\left(\mathbf{A}\right)} \mbox{''}
\]
as it is supposed to define the partition function of the $\mathrm{U}(1)^{n}$ Chern-Simons theory.

Although is it extensively used in the context of Quantum Field Theory (QFT), it is well-known that such a functional integral is in general mathematically ill-defined. Our main goal is to show that we can extract a well-defined quantity from this functional integral in the same spirit as it is possible to extract physical numbers from infinite integrals in the perturbative approach of QFT. This finite quantity will be our partition function. Of course, this extraction must rely on some mathematically consistent procedure, as is renormalisation in perturbative QFT.  This procedure is derived from exact sequence \eqref{SES_1} and more specifically from decomposition \eqref{decompA}.

Decomposing each component $A_{i}\in H^1_{\mathrm{DB}}(M)$ of $\mathbf{A}\in\left(H^1_{\mathrm{DB}}(M)\right)^{n}$ as such we get:
\begin{equation}
\label{decompboldA}
\mathbf{A} 
= \mathbf{A}_{\mathbf{m}_{\mathbf{A}}} 
+ \mathbf{A}_{\boldsymbol{\kappa}_{\mathbf{A}}} 
+ \boldsymbol{\alpha}_{\perp} 
+ \boldsymbol{\alpha}_{0}   
\end{equation}
where $\mathbf{m}_{\mathbf{A}}\in\left(FH^{2}(M)\right)^{n}\cong\left(\Z^{b_{1}}\right)^{n}$, $\boldsymbol{\kappa}_{\mathbf{A}}\in\left(TH^{2}(M)\right)^{n}$, $\boldsymbol{\alpha}_{\perp}\in\left(\slfrac{\Omega^{1}(M)}{\Omega^{1}_{\mathrm{cl}}(M)}\right)^{n}$ and $\boldsymbol{\alpha}_{0}\in\left(\slfrac{\Omega^{1}_\mathrm{cl}(M)}{\Omega^{1}_{\Z}(M)}\right)^{n}\cong\left(\left(\slfrac{\R}{\Z}\right)^{b_{1}}\right)^{n}$.

According to decomposition \eqref{decompboldA}, we get that
\begin{align*}
\mbox{``} \mathcal{Z}_{\mathrm{CS}_{\mathbf{C}}}
= &\frac{1}{\mathcal{N}_{\mathrm{CS}_{\mathbf{C}}}(M)}
\sum_{\boldsymbol{\kappa}_{\mathbf{A}}\in\left(TH^{2}(M)\right)^{n}}
e^{2\pi i\int_{M}{}^{t}\mathbf{A}_{\boldsymbol{\kappa}_{\mathbf{A}}}\star\mathbf{C}\mathbf{A}_{\boldsymbol{\kappa}_{\mathbf{A}}}}\\
&\sum_{\mathbf{m}_{\mathbf{A}}\in\left(FH^{2}(M)\right)^{n}}e^{2\pi i\left(\int_{M}{}^{t}\mathbf{A}_{\mathbf{m}_{\mathbf{A}}}\star\mathbf{C}\mathbf{A}_{\mathbf{m}_{\mathbf{A}}}
+ \int_{M}{}^{t}\mathbf{A}_{\boldsymbol{\kappa}_{\mathbf{A}}}\star\mathbf{C}\mathbf{A}_{\mathbf{m}_{\mathbf{A}}}+ \int_{M}{}^{t}\mathbf{A}_{\mathbf{m}_{\mathbf{A}}}\star\mathbf{C}\mathbf{A}_{\boldsymbol{\kappa}_{\mathbf{A}}}\right)}\\
&\int_{\left(\slfrac{\Omega^{1}_{\mathrm{cl}}(M)}{\Omega^{1}_{\Z}(M)}\right)^{n}}
\,\mathscr{D}\boldsymbol{\alpha}_{0}\,
e^{2\pi i\left(\int_{M}{}^{t}\mathbf{A}_{\mathbf{m}_{\mathbf{A}}}\star\mathbf{C}\boldsymbol{\alpha}_{0}
+ \int_{M}{}^{t}\boldsymbol{\alpha}_{0}\star\mathbf{C}\mathbf{A}_{\mathbf{m}_{\mathbf{A}}}\right)}\\
&\int_{\left(\slfrac{\Omega^{1}(M)}{\Omega^{1}_{\mathrm{cl}}(M)}\right)^{n}}
\,\mathscr{D}\boldsymbol{\alpha}_{\perp}\,
e^{2\pi i\left(\int_{M}{}^{t}\mathbf{A}_{\mathbf{m}_{\mathbf{A}}}\star\mathbf{C}\boldsymbol{\alpha}_{\perp} 
+ \int_{M}{}^{t}\boldsymbol{\alpha}_{\perp}\star\mathbf{C}\mathbf{A}_{\mathbf{m}_{\mathbf{A}}} 
+ \int_{M}{}^{t}\boldsymbol{\alpha}_{\perp}\star\mathbf{C}\boldsymbol{\alpha}_{\perp}\right)}
\mbox{''}
\end{align*}

Using the fact that
\[
\int_{M}{}^{t}\mathbf{A}_{\mathbf{m}_{\mathbf{A}}}\star\mathbf{C}\boldsymbol{\alpha}_{0}
\underset{\Z}{=} {}^{t}\mathbf{m}_{\mathbf{A}}\mathbf{C}\boldsymbol{\theta}_{\mathbf{A}}
\]
and
\[
\int_{M}{}^{t}\boldsymbol{\alpha}_{0}\star\mathbf{C}\mathbf{A}_{\mathbf{m}_{\mathbf{A}}}
\underset{\Z}{=} {}^{t}\boldsymbol{\theta}_{\mathbf{A}}\mathbf{C}\mathbf{m}_{\mathbf{A}}
\]
where $\boldsymbol{\theta}_{\mathbf{A}} = \left(\boldsymbol{\theta}_{A_{1}},\hdots,\boldsymbol{\theta}_{A_{n}}\right)\in\left(\left(\slfrac{\R}{\Z}\right)^{b_{1}}\right)^{n}$, we can write
\begin{align*}
e^{2\pi i\left(\int_{M}{}^{t}\mathbf{A}_{\mathbf{m}_{\mathbf{A}}}\star\mathbf{C}\boldsymbol{\alpha}_{0}
+ \int_{M}{}^{t}\boldsymbol{\alpha}_{0}\star\mathbf{C}\mathbf{A}_{\mathbf{m}_{\mathbf{A}}}\right)}
=& e^{2\pi i\left({}^{t}\mathbf{m}_{\mathbf{A}}\mathbf{C}\boldsymbol{\theta}_{\mathbf{A}}
+ {}^{t}\boldsymbol{\theta}_{\mathbf{A}}\mathbf{C}\mathbf{m}_{\mathbf{A}}\right)}\\
=& e^{2\pi i\left({}^{t}\mathbf{m}_{\mathbf{A}}\mathbf{C}\boldsymbol{\theta}_{\mathbf{A}}
+ {}^{t}\left({}^{t}\mathbf{m}_{\mathbf{A}}{}^{t}\mathbf{C}\boldsymbol{\theta}_{\mathbf{A}}\right)\right)}\\
=& e^{2\pi i\left({}^{t}\mathbf{m}_{\mathbf{A}}\mathbf{C}\boldsymbol{\theta}_{\mathbf{A}}
+ {}^{t}\mathbf{m}_{\mathbf{A}}{}^{t}\mathbf{C}\boldsymbol{\theta}_{\mathbf{A}}\right)}\\
=& e^{2\pi i\left({}^{t}\mathbf{m}_{\mathbf{A}}\left(\mathbf{C}+{}^{t}\mathbf{C}\right)\boldsymbol{\theta}_{\mathbf{A}}\right)}\\
e^{2\pi i\left(\int_{M}{}^{t}\mathbf{A}_{\mathbf{m}_{\mathbf{A}}}\star\mathbf{C}\boldsymbol{\alpha}_{0}
+ \int_{M}{}^{t}\boldsymbol{\alpha}_{0}\star\mathbf{C}\mathbf{A}_{\mathbf{m}_{\mathbf{A}}}\right)}
=& e^{2\pi i\left({}^{t}\mathbf{m}_{\mathbf{A}}\mathbf{K}\boldsymbol{\theta}_{\mathbf{A}}\right)}
\end{align*}
where
\begin{equation}
\mathbf{K} = \mathbf{C}+{}^{t}\mathbf{C}\in\mathrm{Sym}_{n}(\Z). 
\end{equation}
Note that $\mathbf{K}$ has even integers on the diagonal and $n(n+1)/2$ independent entries, the number of independent coupling constant thus remaining the same. Hence, we can write
\begin{align*}
\int_{\left(\slfrac{\Omega^{1}_{\mathrm{cl}}(M)}{\Omega^{1}_{\Z}(M)}\right)^{n}}
\,\mathscr{D}\boldsymbol{\alpha}_{0}\,
e^{2\pi i\left(\int_{M}{}^{t}\mathbf{A}_{\mathbf{m}_{\mathbf{A}}}\star\mathbf{C}\boldsymbol{\alpha}_{0}
+ \int_{M}{}^{t}\boldsymbol{\alpha}_{0}\star\mathbf{C}\mathbf{A}_{\mathbf{m}_{\mathbf{A}}}\right)}
=& \int_{\left(\slfrac{\R}{\Z}\right)^{n}}
\,d\boldsymbol{\theta}_{\mathbf{A}}\,
e^{2\pi i\left({}^{t}\mathbf{m}_{\mathbf{A}}\mathbf{K}\boldsymbol{\theta}_{\mathbf{A}}\right)}\\
=& \; \delta_{\mathbf{K}\mathbf{m}_{\mathbf{A}}}
\end{align*}
where $\delta_{\mathbf{K}\mathbf{m}_{\mathbf{A}}}$ is a regular Kronecker symbol. Importantly, note that $\mathbf{m}_{\mathbf{A}} = \left(\mathbf{m}_{A_{1}},\hdots,\mathbf{m}_{A_{n}}\right)$ where $\forall\,i\in\left\lbrace 1,\hdots, n\right\rbrace,\,\mathbf{m}_{A_{i}}\in\Z^{b_{1}}$, i.e., $\mathbf{m}_{A_{i}} = \left(\left(\mathbf{m}_{A_{i}}\right)_{1},\hdots,\left(\mathbf{m}_{A_{i}}\right)_{b_{1}}\right)$, so that $\mathbf{m}_{\mathbf{A}} = \left(\mathbf{m}_{A_{1}},\hdots,\mathbf{m}_{A_{n}}\right) = \left(\left(\mathbf{m}_{A_{1}}\right)_{1},\hdots,\left(\mathbf{m}_{A_{1}}\right)_{b_{1}},\hdots,\left(\mathbf{m}_{A_{n}}\right)_{1},\hdots,\left(\mathbf{m}_{A_{n}}\right)_{b_{1}}\right)$, and 
\[
\mathbf{K}\mathbf{m}_{\mathbf{A}}
= \begin{pmatrix}
\sum\limits_{i=1}^{n}k_{1i}\mathbf{m}_{A_{i}}\\
\vdots\\
\sum\limits_{i=1}^{n}k_{ni}\mathbf{m}_{A_{i}}
\end{pmatrix}
\]
each line of the column vector being itself a column vector. The way we interpret the Kronecker symbol of such object is
\[
\delta_{\mathbf{K}\mathbf{m}_{\mathbf{A}}}
= \delta_{\sum\limits_{i=1}^{n}k_{1i}\mathbf{m}_{A_{i}}}\hdots\delta_{\sum\limits_{i=1}^{n}k_{ni}\mathbf{m}_{A_{i}}}.
\]

We will assume first that $\mathbf{K}$ is nondegenerate, so that 
\begin{align*}
\delta_{\mathbf{K}\mathbf{m}_{\mathbf{A}}} 
=& \; \delta_{\mathbf{m}_{\mathbf{A}}}\\
=& \; \delta_{\mathbf{m}_{A_{1}}}\hdots\delta_{\mathbf{m}_{A_{n}}}\\
=& \; \delta_{\left(\mathbf{m}_{A_{1}}\right)_{1}}\hdots\delta_{\left(\mathbf{m}_{A_{1}}\right)_{b_{1}}}
\hdots
\delta_{\left(\mathbf{m}_{A_{n}}\right)_{1}}\hdots\delta_{\left(\mathbf{m}_{A_{n}}\right)_{b_{1}}}    
\end{align*}

Moreover, by using the fact that 
\[
\int_{M}{}^{t}\mathbf{A}_{\boldsymbol{\kappa}_{\mathbf{A}}}\star\mathbf{C}\mathbf{A}_{\boldsymbol{\kappa}_{\mathbf{A}}}
\underset{\Z}{=} -{}^{t}\boldsymbol{\kappa}_{\mathbf{A}}\left(\mathbf{C}\otimes\mathbf{Q}\right)\boldsymbol{\kappa}_{\mathbf{A}}
\]

we obtain the following expression of the partition function
\begin{align*}
\mbox{``} \mathcal{Z}_{\mathrm{CS}_{\mathbf{C}}}
= &\frac{1}{\mathcal{N}_{\mathrm{CS}_{\mathbf{C}}}(M)}
\sum_{\boldsymbol{\kappa}_{\mathbf{A}}\in\left(TH^{2}(M)\right)^{n}}
e^{-2\pi i{}^{t}\boldsymbol{\kappa}_{\mathbf{A}}\left(\mathbf{C}\otimes\mathbf{Q}\right)\boldsymbol{\kappa}_{\mathbf{A}}}\\
&\sum_{\mathbf{m}_{\mathbf{A}}\in\left(FH^{2}(M)\right)^{n}}
\delta_{\mathbf{K}\mathbf{m}_{\mathbf{A}}}e^{2\pi i\left(\int_{M}{}^{t}\mathbf{A}_{\mathbf{m}_{\mathbf{A}}}\star\mathbf{C}\mathbf{A}_{\mathbf{m}_{\mathbf{A}}}
+ \int_{M}{}^{t}\mathbf{A}_{\boldsymbol{\kappa}_{\mathbf{A}}}\star\mathbf{C}\mathbf{A}_{\mathbf{m}_{\mathbf{A}}}+ \int_{M}{}^{t}\mathbf{A}_{\mathbf{m}_{\mathbf{A}}}\star\mathbf{C}\mathbf{A}_{\boldsymbol{\kappa}_{\mathbf{A}}}\right)}\\
&\int_{\left(\slfrac{\Omega^{1}(M)}{\Omega^{1}_{\mathrm{cl}}(M)}\right)^{n}}
\mathscr{D}\boldsymbol{\alpha}_{\perp}
e^{2\pi i\left(\int_{M}{}^{t}\mathbf{A}_{\mathbf{m}_{\mathbf{A}}}\star\mathbf{C}\boldsymbol{\alpha}_{\perp} 
+ \int_{M}{}^{t}\boldsymbol{\alpha}_{\perp}\star\mathbf{C}\mathbf{A}_{\mathbf{m}_{\mathbf{A}}} 
+ \int_{M}{}^{t}\boldsymbol{\alpha}_{\perp}\star\mathbf{C}\boldsymbol{\alpha}_{\perp}\right)}
\mbox{''}
\end{align*}
The next step is to sum over the topological sector $\left(FH^{2}(M)\right)^{n}$, so that, taking into account the previous discussion about $\delta_{\mathbf{K}\mathbf{m}_{\mathbf{A}}}$, we get
\begin{align*}
\mbox{``} \mathcal{Z}_{\mathrm{CS}_{\mathbf{C}}}
= \frac{1}{\mathcal{N}_{\mathrm{CS}_{\mathbf{C}}}(M)}
&\sum_{\boldsymbol{\kappa}_{\mathbf{A}}\in\left(TH^{2}(M)\right)^{n}}
e^{-2\pi i{}^{t}\boldsymbol{\kappa}_{\mathbf{A}}\left(\mathbf{C}\otimes\mathbf{Q}\right)\boldsymbol{\kappa}_{\mathbf{A}}}\\
&\int_{\left(\slfrac{\Omega^{1}(M)}{\Omega^{1}_{\mathrm{cl}}(M)}\right)^{n}}
\mathscr{D}\boldsymbol{\alpha}_{\perp}
e^{2\pi i\int_{M}{}^{t}\boldsymbol{\alpha}_{\perp}\star\mathbf{C}\boldsymbol{\alpha}_{\perp}}
\mbox{''}
\end{align*}

We observe here a convenient full decoupling of the two remaining topological sectors, $\left(\slfrac{\Omega^{1}(M)}{\Omega^{1}_{\mathrm{cl}}(M)}\right)^{n}$ and $\left(TH^{2}(M)\right)^{n}$, exactly like in the standard $\mathrm{U}(1)$ case \cite{GT2014}. The contribution of the topological sector $\left(\slfrac{\Omega^{1}(M)}{\Omega^{1}_{\mathrm{cl}}(M)}\right)^{n}$ is infinite dimensional, and we choose to eliminate it\footnote{A lot of authors \cite{Sch1978,BBRT1991, AS1991} extract from this part the Reidemeister torsion of $M$. However, they do not work with the gauge classes of fields. They fix the gauge, and for that they usually introduce a metric, which they want afterwards to get rid of, as the theory is expected to be topological, i.e., the partition function (and the expectation values of observables) should not depend on any metric.} thanks to the normalization, i.e., by writing formally
\begin{align}
\mbox{``}\,\,&\mathcal{N}_{\mathrm{CS}_{\mathbf{C}}}(M)
= \int_{\left(\slfrac{\Omega^{1}(M)}{\Omega^{1}_{\mathrm{cl}}(M)}\right)^{n}}
\hspace{-0.25cm}\mathscr{D}\boldsymbol{\alpha}_{\perp}
e^{2\pi i\int_{M}{}^{t}\boldsymbol{\alpha}_{\perp}\star\mathbf{C}\boldsymbol{\alpha}_{\perp}}.
\,\,\mbox{''}
\label{Normalization}
\end{align}
Therefore, only the contribution of the sector $\left(TH^{2}(M)\right)^{n}$ remains, which we will regard as our \textit{definition} of the partition function $\mathcal{Z}_{\mathrm{CS}_{\mathbf{C}}}(M)$:
\begin{defi}
In the following,
\begin{equation}
\mathcal{Z}_{\mathrm{CS}_{\mathbf{C}}}
:= \sum_{\boldsymbol{\kappa}_{\mathbf{A}}\in\left(TH^{2}(M)\right)^{n}}
e^{-2\pi i{}^{t}\boldsymbol{\kappa}_{\mathbf{A}}\left(\mathbf{C}\otimes\mathbf{Q}\right)\boldsymbol{\kappa}_{\mathbf{A}}}.
\end{equation}
\end{defi}

In this expression of the partition function, the quantity ${}^{t}\boldsymbol{\kappa}_{\mathbf{A}} \left(\mathbf{C} \otimes \mathbf{Q}\right) \boldsymbol{\kappa}_{\mathbf{A}}$ is defined in $\slfrac{\R}{\Z}$ with $\boldsymbol{\kappa}_{\mathbf{A}}$ having $n$ components $\boldsymbol{\kappa}_{A_i} \in TH^2(M)$. Let us pick up a representative $\vec{\kappa}_{A_i} \in \Z^{r}$ for each $\boldsymbol{\kappa}_{A_i} \in \slfrac{\Z^{r}}{\mathbf{L}_{0}\Z^{r}}$. We write $\vec{\kappa}_{\mathbf{A}}$ the corresponding representative of of $\boldsymbol{\kappa}_{\mathbf{A}}$. Then, as $Q$ is symmetric, we have
\begin{equation} \label{FromMtoK}
{}^{t}\vec{\kappa}_{\mathbf{A}} \left(\mathbf{C} \otimes \mathbf{Q}\right) \vec{\kappa}_{\mathbf{A}}
= \frac{1}{2}{}^{t}\vec{\kappa}_{\mathbf{A}} \left(\mathbf{K} \otimes \mathbf{Q}\right) \vec{\kappa}_{\mathbf{A}} ,
\end{equation}
with $\mathbf{K} = \mathbf{C} + {}^{t}\mathbf{C}$. It is not difficult to check that a different choice of representatives $\vec{\kappa}_{A_i}$ changes the left and right hand side of the above equality by an integer. From now on, we adopt the convention that in the evaluation of the partition function, we will use a given representative for each class in $TH^2(M)$, so we can write
\begin{equation}
\mathcal{Z}_{\mathrm{CS}_{\mathbf{C}}}
= \sum_{\boldsymbol{\kappa}_{\mathbf{A}}\in\left(TH^{2}(M)\right)^{n}}
e^{-\pi i {}^{t}\boldsymbol{\kappa}_{\mathbf{A}} \left(\mathbf{K} \otimes \mathbf{Q}\right) \boldsymbol{\kappa}_{\mathbf{A}}} .
\end{equation}
Alternatively, we can consider as fundamental representatives of the elements of $\Z_{p_1} \oplus \cdots \oplus \Z_{p_t} \cong \slfrac{\Z^r}{\mathbf{L}_0\Z^r}$ the $t$-uples of integers $\left(u_1, \cdots , u_t\right)$ with $u_i \in \left[\kern-0.15em\left[ {0,p_i-1} \right]\kern-0.15em\right]$. We consider the subset $\left[\kern-0.15em\left[ {\vec{0},\vec{p}-1} \right]\kern-0.15em\right] := \left[\kern-0.15em\left[ {0,p_1-1} \right]\kern-0.15em\right] \times \cdots \times \left[\kern-0.15em\left[ {0,p_t-1} \right]\kern-0.15em\right]$ of $\N^t$ the elements of which are written $\vec{u}$. The linking form $\mathbf{Q}$ is then represented by a $t \times t$ rational symmetric and invertible matrix, still denoted by $\mathbf{Q}$, which acts on $\left[\kern-0.15em\left[ {\vec{0},\vec{p}-1} \right]\kern-0.15em\right]$\footnote{An explicit construction of the $t \times t$ matrix $\mathbf{Q}$, starting from $\mathbf{L}_0^{-1}$ can be found in Chapter 4 of \cite{TM2023}}. The elements of the Cartesian product $\left[\kern-0.15em\left[ {\vec{0},\vec{p}-1} \right]\kern-0.15em\right]^n$ are obviously representatives of $\left(TH^{2}(M)\right)^{n}$, and if we denote by $\boldsymbol{u}$ these elements then the partition function takes the form
\begin{equation}
\mathcal{Z}_{\mathrm{CS}_{\mathbf{C}}}
= \sum_{\boldsymbol{u} \in \left[\kern-0.15em\left[ {\vec{0},\vec{p}-1} \right]\kern-0.15em\right]^n}
e^{-\pi i {}^{t}\boldsymbol{u} \left(\mathbf{K}\otimes\mathbf{Q}\right) \boldsymbol{u}} .
\end{equation}
From now on, and whichever point of view is adopted, we use the convention that in the evaluation of ${}^{t}\boldsymbol{\kappa}_{\mathbf{A}} \left(\mathbf{K} \otimes \mathbf{Q}\right) \boldsymbol{\kappa}_{\mathbf{A}}$ the same representatives are used for $\boldsymbol{\kappa}_{\mathbf{A}}$ and ${}^{t}\boldsymbol{\kappa}_{\mathbf{A}}$.

\begin{rema}
In the case where $\mathbf{K}$ is degenerate, there are infinitely many solutions to the equation
\[
\mathbf{K}\mathbf{m}_{\mathbf{A}} = 0
\]
but the contribution of each solution to the functional integral is the same. Thus, we can ``factor'' the cardinality of the set of solutions (which is infinite), i.e., the cardinality of the kernel of $\mathbf{K}$, and eliminate it with the normalization, in such a way that the above definition of the partition function still holds. 
\end{rema}

\begin{rema}[\cite{DW1990, Freed1995, Freed2002}]
For a given group $G$, the Chern-Simons theories with gauge group $G$ are classified by $H^{4}\left(BG\right)$, where $BG$ is the classifying space of $G$. 

For $G = \mathrm{U}(1)$, we have 
\[
H^{*}\left(B\mathrm{U}(1)\right)\cong\Z\left[t\right],
\]
the ring of polynomials with $\Z$ coefficients in a variable $t$ which is of degree $2$, so
\[
H^{4}\left(B\mathrm{U}(1)\right)\cong\Z,
\]
which is indeed the space to which the coupling constant belongs for the $\mathrm{U}(1)$ Chern-Simons theory. 

For $G = \mathrm{U}(1)^{n}$, we have \cite{Tod1986} 
\[
H^{*}\left(B\left(\mathrm{U}(1)^{n}\right)\right) = \Z\left[t_{1},\hdots,t_{n}\right],
\]
the ring of polynomials with $\Z$ coefficients in $n$ variables $t_{i}$ which are of degree $2$, so
\[
H^{4}\left(B\left(\mathrm{U}(1)^{n}\right)\right) = \Z\left[t_{1},\hdots,t_{n}\right]_{2},
\]
the $\Z$-module of homogeneous polynomials of degree $2$ in $n$ variables $t_{i}$ which are of degree $2$. This $\Z$-module is $\left(\frac{n\left(n+1\right)}{2}\right)$-dimensional (an obvious basis being $\left\lbrace t_{i}t_{j}\,\vert\, 1\leq i\leq j\leq n\right\rbrace$). This is indeed the space to which the coupling constant belongs for the $\mathrm{U}(1)^{n}$ Chern-Simons theory as shown in remark \ref{BFinCS}.
\end{rema}

\subsection{The reciprocity formula and CS-duality}

\begin{thm}[\cite{Tur1998}]
Let $M$ be a closed oriented $3$-manifold obtained from $S^{3}$ by \textbf{integral} surgery along a link $\boldsymbol{\mathcal{L}}\in S^{3}$ whose associated linking matrix $\mathbf{L}=\mathbf{L}_{\boldsymbol{\mathcal{L}}}\in\mathrm{Sym}_{m}(\Z)$ can be written as
\[
\mathbf{L}
=\begin{pmatrix}
\mathbf{L}_{0} & \boldsymbol{0} \\
\boldsymbol{0} & \boldsymbol{0}
\end{pmatrix}\in\mathrm{Sym}_{m}(\Z)
\]
where $\mathbf{L}_{0}\in\mathrm{Sym}_{r}(\Z)$ has nonzero determinant. Consider also an even integral symmetric matrix (i.e., a symmetric integral matrix with even numbers on the diagonal) $\mathbf{K}\in\mathrm{Sym}_{n}(\Z)$ which can be written as
\[
\mathbf{K}
=\begin{pmatrix}
\mathbf{K}_{0} & \boldsymbol{0} \\
\boldsymbol{0} & \boldsymbol{0}
\end{pmatrix}\in\mathrm{Sym}_{n}(\Z)
\]
where $\mathbf{K}_{0}\in\mathrm{Sym}_{s}(\Z)$ has nonzero determinant. Then
\begin{align}
\label{eq:rec3} 
\nonumber
&\frac{1}{\abs{\determinant{\mathbf{K}_{0}}}^{m-\frac{r}{2}}}
\sum\limits_{\mathbf{x}_{\boldsymbol{\mathcal{L}}}\in\left(\slfrac{\Z^{s}}{\mathbf{K}_{0}\Z^{s}}\right)^{m}}
e^{\pi i{}^{t}\mathbf{x}_{\boldsymbol{\mathcal{L}}}
\left(\mathbf{L}\otimes\mathbf{K}_{0}^{-1}\right)
\mathbf{x}_{\boldsymbol{\mathcal{L}}}}\\ 
&=\frac{1}{\abs{\determinant{\mathbf{L}_{0}}}^{n-\frac{s}{2}}}
e^{\frac{i\pi}{4}\sigma\left(\mathbf{K}\right)\sigma\left(\mathbf{L}\right)}
\sum_{\boldsymbol{\kappa}_{\mathbf{A}}\in\left(\slfrac{\Z^{r}}{\mathbf{L}_{0}\Z^{r}}\right)^{n}}
e^{-\pi i {}^{t}\boldsymbol{\kappa}_{\mathbf{A}}
\left(\mathbf{K}\otimes\mathbf{L}^{-1}_{0}\right)
\boldsymbol{\kappa}_{\mathbf{A}}}  
\end{align}
\end{thm}

\begin{rema}
Recall that, under this form, the RHS makes sense when choosing the same representative of $\boldsymbol{\kappa}_{\mathbf{A}}$ on both sides of $\left(\mathbf{K}\otimes\mathbf{L}^{-1}_{0}\right)$, the same is true for the LHS.
\end{rema}
\begin{rema}\label{Inv_under_Kirby}  The symmetry of the action in definition \ref{def_of_C_matrix} now reads $(K,L) \to (-K,-L)$ and both sides of the reciprocity formula are invariant under that symmetry. Furthermore, the left-hand side of the equation is invariant when performing the second Kirby move on $\mathbf{L}$ and it only changes by a phase (which is compensated on the right-hand side) when performing the first Kirby move \cite{TM2023}.
\end{rema}
When carefully examining relation \eqref{eq:rec3}  where the sum appearing in the right-hand side is a partition function, it seems natural to wonder whether the sum in the left-hand-side can also be a partition function. To check if this is true, let us start by recalling that any closed oriented and smooth 3-manifold can be obtained by an even surgery in $S^3$ \cite{Sav2012}. Thus, we can restrict the linking matrices that we have considered so far to be even. This way, the matrices $\mathbf{L}$ and $\mathbf{K}$ are both integral, even and symmetric. Then, on the one hand we can associate to $\mathbf{L}$ an upper triangular matrix $\mathbf{\Hat{{C}}}$ by setting 
\[
\Hat{C}_{ij}
= \begin{cases} 
      L_{ij} & \mbox{ if } i < j \\
      \frac{1}{2}L_{ii}  & \mbox{ if } i=j \\
      0 & \mbox{ otherwise,}
\end{cases}
\]
and on the other hand, since the integral matrix $\mathbf{K}$ is even and symmetric, it can be seen as the linking matrix of some even surgery in $S^3$. Let $M_K$ be a manifold obtained by such a surgery. Of course $M_K$ is not unique since the linking matrix $\mathbf{K}$ just defines the homology of $M_K$ and a linking form, $\mathbf{K}_0^{-1}$, on its torsion sector $\slfrac{\Z^{s}}{\mathbf{K}_{0}\Z^{s}}$. Now, let us consider the following $\mathrm{U}(1)^m$ Chen-Simons action
\begin{equation}
\hat{S}_{CS}(M_K) = \int_{M_K} {}^{t}\mathbf{A} \star \mathbf{\Hat{C}} \mathbf{A} ,
\end{equation}
where $\mathbf{A}$ is a column vector whose components are elements of $H_{DB}^1(M_K)$. From what we have done before, we straightforwardly deduce the corresponding partition function 
\begin{equation}
\mathcal{Z}_{\mathrm{CS}_{\mathbf{M_K}}}
= \sum\limits_{\mathbf{x}\in\left(\slfrac{\Z^{s}}{\mathbf{K}_{0}\Z^{s}}\right)^{m}}
e^{- 2 \pi i {}^{t}\mathbf{x} 
\left(\mathbf{\Hat{C}} \otimes \mathbf{K}_{0}^{-1}\right)
\mathbf{x}} ,
\end{equation}
and with the convention that we use the same representative in the evaluation of ${}^{t}\mathbf{x} 
\left(\mathbf{\Hat{C}} \otimes \mathbf{K}_{0}^{-1}\right)
\mathbf{x}$, we finally obtain
\begin{equation}
\mathcal{Z}_{\mathrm{CS}_{\mathbf{M_K}}}
= \sum\limits_{\mathbf{x}\in\left(\slfrac{\Z^{s}}{\mathbf{K}_{0}\Z^{s}}\right)^{m}}
e^{ - \pi i {}^{t}\mathbf{x} 
\left(\mathbf{L} \otimes \mathbf{K}_{0}^{-1}\right)
\mathbf{x}} .
\end{equation}
Up to complex conjugation, this is obviously the sum appearing in the left-hand side of \eqref{eq:rec3}. This yields a duality property as stated in the following corollary.

\begin{coro}
    To any $\mathrm{U}(1)^n$ Chern-Simons theory with $n \times n$ even symmetric coupling matrix $\mathbf{K}$ and $m \times m$ even linking matrix $\mathbf{L}$ we can associate a $\mathrm{U}(1)^m$ Chern-Simons theory with an even symmetric coupling matrix $-\mathbf{L}$ and an even linking matrix $\mathbf{K}$. We refer to this as a CS-duality of $\mathrm{U}(1)^{n}$ Chern-Simons theories on closed oriented smooth $3$-manifolds with each other.
\end{coro}

\begin{exple}
The above Corollary implies that to a $\mathrm{U}(1)$ Chern-Simons theory of a $3$-manifold with even linking matrix $\mathbf{L}$, as studied in \cite{GT2014}, is associated a $\mathrm{U}(1)^m$ Chern-Simons theory with coupling matrix $\mathbf{L}$ and linking matrix $\mathbf{K}=(-2k)$. This linking matrix can be seen as the one (linking matrix) of an even surgery in $S^3$ of the lens space $L(2k,1)$ for instance, the linking form being $\mathbf{K}_0^{-1}=\mathbf{K}^{-1}=(-1/2k)$.
\end{exple}

\begin{exple}
    Partition function for the gauge group $\mathrm{U}(1)^{2}$ of the lens space $L(2,1)$ with the coupling matrix $\mathbf{K} = \begin{pmatrix}
        2 & 1\\
        1 & 4
    \end{pmatrix}$
\begin{align*}
    \mathcal{Z}_{CS}&= \sum_{x \in (\mathbb{Z}_2)^2} \exp \left(-\pi i \hspace{0.1cm}{}^{t}x (\mathbf{K} \otimes \frac{-1}{2}) x \right)\\
    &= \exp \left( \pi i \frac{1}{2} \begin{pmatrix}
        0 & 0 
    \end{pmatrix}\begin{pmatrix}
        2 & 1\\
        1& 4\\
    \end{pmatrix} \begin{pmatrix}
        0 \\
        0 
    \end{pmatrix}\right)+ \exp \left( \pi i \frac{1}{2} \begin{pmatrix}
        1 & 0 
    \end{pmatrix}\begin{pmatrix}
        2 & 1\\
        1& 4\\
    \end{pmatrix} \begin{pmatrix}
        1 \\
        0 
    \end{pmatrix}\right)\\
    &+ \exp \left( \pi i \frac{1}{2} \begin{pmatrix}
        0 & 1 
    \end{pmatrix}\begin{pmatrix}
        2 & 1\\
        1& 4\\
    \end{pmatrix} \begin{pmatrix}
        0 \\
        1 
    \end{pmatrix}\right)+ \exp \left(- \pi i \frac{1}{2} \begin{pmatrix}
        1 & 1 
    \end{pmatrix}\begin{pmatrix}
        2 & 1\\
        1& 4\\
    \end{pmatrix} \begin{pmatrix}
        1 \\
        1 
    \end{pmatrix}\right)\\
    &= 1 + \exp \left( -\pi i \frac{1}{2} 2
    \right)+ \exp \left( -\pi i \frac{1}{2} 4
    \right)+ 
    \exp \left( -\pi i \frac{1}{2} 8
    \right) \\
    &= 1 -1 + 1 + 1 = 2
\end{align*}
To be compared with
\begin{align*}
    \sum_{y \in \mathbb{Z}^2/\mathbf{K}\mathbb{Z}^2} \exp \left(  \pi i  \hspace{0.1cm}{}^{t}y (2 \otimes \mathbf{K}^{-1}) y \right)
\end{align*}
Note that $\mathrm{det}(\mathbf{K}) = 7$ and since it is a prime number, $ \mathbb{Z}^2/\mathbf{K}\mathbb{Z}^2 \simeq \mathbb{Z}_7$. Thus, every vector of the lattice $ \mathbb{Z}^2/\mathbf{K}\mathbb{Z}^2 $ is a generator of the group. So we can choose the vector $\begin{pmatrix}
    1\\0
\end{pmatrix}$ to be our generator. And so we have

\begin{align*}
        &\sum_{y \in \mathbb{Z}^2/\mathbf{K}\mathbb{Z}^2} \exp \left(  \pi i  \hspace{0.1cm}{}^{t}y (2 \otimes \mathbf{K}^{-1}) y \right) =     \sum_{n \in \mathbb{Z}_7} \exp \left(2  \pi i  n \begin{pmatrix}
            1 &0
        \end{pmatrix} \begin{pmatrix}
            4/7 & -1/7\\
            -1/7 & 2/7
        \end{pmatrix} \begin{pmatrix}
            1\\0
        \end{pmatrix} n\right)\\
         &\sum_{n \in \mathbb{Z}_7} \exp \left(2  \pi i 
            \frac{4n^2}{7} \right)=1 + e^{ \pi i \frac{8}{7}}+ e^{ \pi i \frac{4}{7}}+ e^{ \pi i \frac{2}{7}}+ e^{ \pi i \frac{2}{7}}+ e^{ \pi i \frac{4}{7}}+ e^{ \pi i \frac{8}{7}}.
\end{align*}
Which is not the easiest calculation but it computes to $i \sqrt{7} $. Now for the reciprocity formula, we would have
\[
\frac{1}{(\mathrm{det}(\mathbf{K}))^{\frac{1}{2}}} i \sqrt{7} = \frac{1}{2} \exp \left( i \pi \frac{1}{4} (\sigma (\mathbf{K})) \right) 2.
\]
Finally, since $\sigma (\mathbf{K}) = 2$, we would have
\[
\frac{1}{7^{\frac{1}{2}}} i \sqrt{7} = \frac{1}{2} \exp \left( i \pi \frac{1}{2}  \right) 2,
\]
which is true.
\end{exple}

\begin{exple}
    Partition function for the gauge group $\mathrm{U}(1)^{2}$ of the lens space $L(p,q)$ with the coupling matrix $\mathbf{K}= \begin{pmatrix}
        0 & k\\
        k& 0\\
    \end{pmatrix}$
    \begin{align*}
          \mathcal{Z}_{CS}&= \sum_{{}^{t}(x_1 , x_2) \in (\mathbb{Z}_p)^2} \exp \left(-\pi i \begin{pmatrix}
              x_1 & x_2
          \end{pmatrix} (\mathbf{K} \otimes \frac{-q}{p}) \begin{pmatrix}
              x_1 \\ x_2
          \end{pmatrix} \right) =\sum_{{}^{t}(x_1 , x_2) \in (\mathbb{Z}_p)^2} \exp \left(2 \pi i 
              x_1   x_2 k
            \frac{q}{p}) 
          \right).
    \end{align*}
    We can perform the sum over $x_2$ first.
    \[
 \sum_{x_2 \in \mathbb{Z}_p} \exp(2i \pi k \frac{x_1 x_2 q}{p}) = \left\{
\begin{array}{ll}
      0 & qk x_1/p  \notin \mathbb{Z} \\
      p & qk x_1/p  \in \mathbb{Z}\\
\end{array}. 
\right. 
\]
And so
     \begin{align*}
          Z_{CS}&= gcd(qk,p) p,
    \end{align*}
and since $p$ and $q$ are co-prime by construction:
         \begin{align*}
          Z_{CS}&= gcd(k,p) p.
    \end{align*}
   This partition function coincides exactly with the partition function of the $\mathrm{U}(1)$ BF theory studied in \cite{MT2016JMP}. It was also shown in the same article that, up to some irrelevant normalisation, this partition function turns out to be the $\Z_k$ Turaev-Viro invariant. Eventually, as explained in \cite{MT2017}, this Turaev-Viro invariant can also be obtained from a Reshetikhin-Turaev construction based on $\Z_k \times \Z_k$. Of course, this holds true for any $3$-manifold and not just for $L(p,q)$ (see Remark \ref{BFinCS}).
\end{exple}

\begin{exple}
We want to show that the partition functions for two homotopy equivalent lens spaces $L(p,q_1)$ and $L(p,q_2)$ are the same (up to complex conjugation). We would have
\[
\sum_{ \vec{x} \in (\mathbb{Z}_p)^n} \exp\left( -2\pi i \frac{q_1}{p}{}^{t}\vec{x}\mathbf{K}\vec{x}\right)
\]
We note that for the lens spaces to be homotopy equivalent, the following must be true
\[
q_1q_2 \equiv \pm \ell^2 \pmod{p}
\]
Note that by construction $q_1$ and $q_2$ are coprime with $p$ but also $\ell$ has to be coprime with $p$ as well since if it was not we would have this, let their common factor be $v$.
\[
q_1 q_2 = \pm\ell^2+vp
\]
for some $v \in \mathbb{Z}$. Then the rhs is divisible by $v$ but the lhs cannot be divisible by $v$ since $v$ is a factor of $p$ and the lhs is coprime to $p$.\\
Now we examine the previous sum and perform the following automorphism on the group: instead of $\vec{x} \in (\mathbb{Z}_p)^n$ we consider $q_2\vec{x}$. It's easy to see that $q_2$ acts as an automorphism of the group $(\mathbb{Z}_p)^n$ since $q_2$ and $p$ are coprime. So now we would have
\begin{align*}
\sum_{ \vec{x} \in (\mathbb{Z}_p)^n} \exp\left( -2\pi i \frac{q_1}{p} \;{}^{t}\vec{x}\mathbf{K}\vec{x}\right)
=& \sum_{ \vec{x} \in (\mathbb{Z}_p)^n} \exp\left(-2\pi i \frac{q_1}{p}q_2^2\;{}^{t}\vec{x}\mathbf{K}\vec{x}\right)\\ 
=& \sum_{ \vec{x} \in (\mathbb{Z}_p)^n} \exp\left( -2\pi i q_2 \frac{q_1q_2}{p}\;{}^{t}\vec{x}\mathbf{K}\vec{x}\right)\\
=& \sum_{ \vec{x} \in (\mathbb{Z}_p)^n} \exp\left( -2\pi i q_2 \frac{\ell^2}{p}\;{}^{t}\vec{x}\mathbf{K}\vec{x}\right)\\
=& \sum_{ \vec{x} \in (\mathbb{Z}_p)^n} \exp\left( -2\pi i  \frac{q_2}{p}\;{}^{t}(\ell\vec{x})\mathbf{K}(\ell\vec{ x})\right)\\
\sum_{ \vec{x} \in (\mathbb{Z}_p)^n} \exp\left( -2\pi i \frac{q_1}{p} \;{}^{t}\vec{x}\mathbf{K}\vec{x}\right)
=& \sum_{ \vec{x} \in (\mathbb{Z}_p)^n} \exp\left( -2\pi i \frac{q_2}{p}\;{}^{t}\vec{x}\mathbf{K}\vec{x}\right)
\end{align*}
Now note that for the last equality we just used that fact that multiplication by $\ell$ is another automorphism of $(\mathbb{Z}_p)^n$. Again for the same reason, since $\ell$ and $p$ are coprime, multiplying each of the $\vec{x}$ by $\ell$ constitutes an automorphism whose partition function is the last sum. 
If we choose $-\ell$ instead, then the result would just be the complex conjugate. The same thing happens if the lens spaces are homeomorphic (i.e. if $
q_1\equiv  \pm q_2  \pmod{p}$ or $q_1 q_2\equiv  \pm 1  \pmod{p}$ ). For the first homeomorphism condition, it is clear that our result is the same for different $q$'s that differ by multiples of $p$ and if they are related by a $-$ sign then the results will just be complex conjugates. The second condition is a sub-case of homotopy equivalence. \\\\

\end{exple}

\section{Conclusion}
The construction and results obtained in this article are presented in the context of closed oriented smooth manifolds of dimension $3$. However, just like the $\mathrm{U}(1)$ Chern-Simons (and BF) theory can be extended to closed oriented smooth $4l+3$-manifolds \cite{GPT2013}, the $\mathrm{U}(1)^n$ theory can also be extended to such $4l+3$-manifolds. In other words, the partition function of a $\mathrm{U}(1)^n$ Chern-Simons depends on algebraic data which are common to all closed oriented smooth $4l+3$-manifolds, these data being a torsion group and a non-degenerate symmetric $\slfrac{\R}{\Z}$-valued bilinear form on it. In the $3$-dimensional case, integral even surgery helps to get these data and provides the necessary ingredient in order to write a reciprocity formula for the partition function. Now, since this reciprocity formula is purely algebraic, it suggests that for any  closed $4l+3$-manifold there should be a linking matrix from which all the necessary data can be obtained. This linking matrix should also be obtainable from some even integral surgery in $S^{4l+3}$, this surgery being defined by a finite set of linked and framed spheres $S^{2l+1}$ in $S^{4l+3}$. Last but not least, the CS-duality we exhibited in the $3$-dimensional case still holds in the general $4l+3$-dimensional case. 

The next step would be to study observables and their expectation values for our $\mathrm{U}(1)^{n}$ Chern-Simons theory. Observables are quite obviously abelian Wilson loops, i.e., $\mathrm{U}(1)$ holonomies, as in the usual $\mathrm{U}(1)$ Chern-Simons theory. The simplest idea is to consider a composition of $\mathrm{U}(1)$ holonomies, one for each DB class composing a field of the $\mathrm{U}(1)^{n}$ Chern-Simons theory. Each oh these fundamental holonomies is defined with the help of a $1$-cycle of $M$. The proceedure yielding the expectation values of such $\mathrm{U}(1)^{n}$ wilson loops should be very similar to that of the $\mathrm{U}(1)$ case \cite{GT2014,MT2016NPB}.

\vskip 1.9 truecm

\noindent {\bf Acknowledgments.}
\vskip 0.5 truecm
P. M. thanks Pr. Alberto Cattaneo for hosting him at UZH during the academic year 2022-2023. He acknowledges partial support of
\begin{itemize}
    \item[-] SNF Grant No. 200020 192080 of the Simons Collaboration on Global Categorical Symmetries,
    \item[-] COST (European Cooperation in Science and Technology, \url{www.cost.eu}) Action 21109 - CaLISTA (Cartan geometry, Lie, Integrable Systems, quantum group Theories for Applications),
    \item[-] NCCR SwissMAP, funded by the Swiss National Science Foundation.
\end{itemize}

M.T. thanks Pr. Vladimir Turaev for answering his questions during his master thesis, which led to the present article.

\bibliographystyle{amsalpha}

\section*{Appendix}
We want to demonstrate how we can turn any linking matrix into a symmetric matrix with even numbers on the diagonal. Assume we have a symmetric linking matrix with at least one odd number on its diagonal. We can always turn the matrix into
\[
\begin{pmatrix}
\mathrm{odd}_1&&&&\\
&\mathrm{odd}_2&&&\\
&&\mathrm{odd}_3&&\\
&&&\ddots&\\
&&&& \mathrm{odd}_k\\
\end{pmatrix}
\]
by performing the second Kirby move from the rows with odd numbers on the diagonal to the even ones. Now pick any row, without loos of generality let us pick the first row. Some components on that row will be even, let's say on column $j$. On those perform the second Kirby move from $j$ to $1$ (for every such $j$) and we will have:
\[
\begin{pmatrix}
n&\mathrm{odd}_{12}&\mathrm{odd}_{13}&\hdots&\mathrm{odd}_{1k}\\
\mathrm{odd}_{21}&\mathrm{odd}_2&&&\\
\mathrm{odd}_{31}&&\mathrm{odd}_3&&\\
\vdots &&&\ddots&\\
\mathrm{odd}_{k1}&&&& \mathrm{odd}_k\\
\end{pmatrix}.
\]
Now we denote the value of the first diagonal component as $n$ (which might not be odd). Note that we keep the same notation for the components of the matrix although the values of those components might have changed. Now we create $n-1$ first Kirby move components
\[
\begin{pmatrix}
-1\\
&\ddots \\
&&-1\\
&&&n&\mathrm{odd}_{12}&\mathrm{odd}_{13}&\hdots&\mathrm{odd}_{1k}\\
&&&\mathrm{odd}_{21}&\mathrm{odd}_2&&&\\
&&&\mathrm{odd}_{31}&&\mathrm{odd}_3&&\\
&&&\vdots &&&\ddots&\\
&&&\mathrm{odd}_{k1}&&&& \mathrm{odd}_k\\
\end{pmatrix}.
\]
From each of those components perform the 2nd Kirby move towards the component with framing $n$.
\[
\begin{pmatrix}
-1&&&-1\\
&\ddots&&\vdots \\
&&-1&-1\\
-1&\hdots&-1&1&\mathrm{odd}_{12}&\mathrm{odd}_{13}&\hdots&\mathrm{odd}_{1k}\\
&&&\mathrm{odd}_{21}&\mathrm{odd}_2&&&\\
&&&\mathrm{odd}_{31}&&\mathrm{odd}_3&&\\
&&&\vdots &&&\ddots&\\
&&&\mathrm{odd}_{k1}&&&& \mathrm{odd}_k\\
\end{pmatrix}
\]
Now isolate the component with framing $+1$ by performing the first Kirby move enough times. First on the top left part (by using the Kirby move once towards each row with $-1$ on the diagonal):
\[
\begin{pmatrix}
-2&&&0\\
&\ddots&&\vdots \\
&&-2&0\\
0&\hdots&0&1&\mathrm{odd}_{12}&\mathrm{odd}_{13}&\hdots&\mathrm{odd}_{1k}\\
&&&\mathrm{odd}_{21}&\mathrm{odd}_2&&&\\
&&&\mathrm{odd}_{31}&&\mathrm{odd}_3&&\\
&&&\vdots &&&\ddots&\\
&&&\mathrm{odd}_{k1}&&&& \mathrm{odd}_k\\
\end{pmatrix}
\]
And then on the bottom right (by performing the Kirby move $\mathrm{odd}_{1j}$ times on each row $j$ or if it is negative perform the inverse of the second Kirby move). Note that by performing the second Kirby move $\mathrm{odd}_{1j}$ times, the parity of the $jj$ component will be even since only the contribution coming from the $+1$ component can change the parity and that contribution happens an odd amount of times. Penultimately, we will have:
\[
\begin{pmatrix}
-2&&&0\\
&\ddots&&\vdots \\
&&-2&0\\
0&\hdots&0&1&0&0&\hdots&0\\
&&&0&\mathrm{even}_2&&&\\
&&&0&&\mathrm{even}_3&&\\
&&&\vdots &&&\ddots&\\
&&&0&&&& \mathrm{even}_k\\
\end{pmatrix}
\]
And finally we will use the inverse of the first Kirby move to remove the "+1" component.
\[
\begin{pmatrix}
-2&&&\\
&\ddots&& \\
&&-2&\\
&&&\mathrm{even}_2&&&\\
&&&&\mathrm{even}_3&&\\
&& &&&\ddots&\\
&&&&&& \mathrm{even}_k\\
\end{pmatrix}.
\]
Now we have an even matrix.
\end{document}